\documentclass[11pt,a4paper]{article}
\pdfoutput=1
\usepackage{jheppub}
\usepackage{amsthm,amsbsy,amsfonts,mathrsfs,enumerate,float,wrapfig,amsmath}
\usepackage[utf8]{inputenc}
\DeclareUnicodeCharacter{2212}{-}
\usepackage{subfigure}

\newcommand{\be}{\begin{equation}}
\newcommand{\ee}{\end{equation}}
\newcommand{\bea}{\begin{eqnarray}}
\newcommand{\eea}{\end{eqnarray}}
\newcommand{\ba}{\begin{align}}
\newcommand{\ea}{\end{align}}

\usepackage{longtable,caption}
\usepackage{amsmath}
\usepackage{amssymb}
\usepackage{MnSymbol}
\DeclareMathOperator{\PE}{PE}

\DeclareMathOperator{\HS}{HS}

\usepackage{tikz}
\usetikzlibrary{decorations.pathmorphing}
\usetikzlibrary{decorations.pathreplacing}
\usetikzlibrary{arrows.meta}
\tikzset{%%
  >={To[length=5pt]}
  }
\usetikzlibrary{shapes, shapes.geometric, shapes.symbols, shapes.arrows, shapes.multipart, shapes.callouts, shapes.misc}
\tikzset{snake it/.style={decorate, decoration=snake}}
\tikzset{7brane/.style={circle, draw=black, fill=black,ultra thick,inner sep=1.5 pt, minimum size=1 pt,}, c/.default={4pt}}
\tikzset{cross/.style={cross out, draw=black,ultra thick, minimum size=2*(#1-\pgflinewidth), inner sep=0pt, outer sep=0pt}, cross/.default={5pt}}
\tikzset{big7brane/.style={circle, draw=black, fill=black,ultra thick,inner sep=2.5 pt, minimum size=1 pt,}, c/.default={4pt}}
\tikzset{u/.style={circle, draw=black, fill=white,inner sep=2 pt, minimum size=2 pt,},f/.style={square, draw=black, fill=white,ultra thick,inner sep=4 pt, minimum size=2 pt,}}

\tikzset{so/.style={circle, draw=black, fill=red,inner sep=2 pt, minimum size=2 pt,},f/.style={square, draw=black, fill=white,ultra thick,inner sep=4 pt, minimum size=2 pt,}}

\tikzset{sp/.style={circle, draw=black, fill=blue,inner sep=2 pt, minimum size=2 pt,},f/.style={square, draw=black, fill=white,ultra thick,inner sep=4 pt, minimum size=2 pt,}}

\tikzset{uf/.style={rectangle, draw=black, fill=white,inner sep=3 pt, minimum size=4 pt,}}

\tikzset{spf/.style={rectangle, draw=black, fill=blue, thick,inner sep=3 pt, minimum size=4 pt, circle, draw=black, fill=blue,thick,inner sep=2 pt, minimum size=2 pt,},f/.style={square, draw=black, fill=white,ultra thick,inner sep=4 pt, minimum size=2 pt,}}

\tikzset{sof/.style={rectangle, draw=black, fill=red, thick,inner sep=3 pt, minimum size=4 pt,}}
\usetikzlibrary{positioning}
\usetikzlibrary{arrows}
\title{Magnetic quivers from brane webs with O7$^+$-planes}
\author[a]{Mohammad Akhond,}
\author[b]{Federico Carta,}
\affiliation[a]{Department of Physics, Swansea University, \\
 Singleton Park, Swansea, SA2 8PP, U.K.}
\affiliation[b]{Department of Mathematical Sciences, Durham University,\\
	Durham, DH$1$ $3$LE, United Kingdom}
\emailAdd{akhondmohammad@gmail.com}
\emailAdd{federico.carta@durham.ac.uk}

\abstract{We consider the Higgs branch of 5d fixed points engineered using brane webs with an O7$^+$-plane. We use the brane construction to propose a set of rules to extract the corresponding magnetic quivers. Such magnetic quivers are generically framed non-simply-laced quivers containing unitary as well as special unitary gauge nodes. We compute the Coulomb branch Hilbert series of the proposed magnetic quivers. In some specific cases, an alternative magnetic quiver can be obtained either using an ordinary brane web or a brane web with an O5-plane. In these cases, we find a match at the level of the Hilbert series.
}
\begin{document}
%\preprint{}
\maketitle
%--------- Section 1: Introduction --------------
\section{Introduction}
One of the most remarkable predictions of string theory is the existence of UV complete quantum field theories in 5 (and 6) spacetime dimensions. This was argued for, long ago, using a combination of considerations in terms of their effective gauge theory description in the IR, their embedding into string/M-theory as the worldvolume theory of branes as well as their holographic description in type IIA supergravity \cite{Seiberg:1996bd,  Intriligator:1997pq,Aharony:1997bh, Douglas:1996xp,Brandhuber:1999np}. Now, over two decades after these initial studies, each of these three avenues has been extensively studied and developed further. Many authors have focused on possible classification of all 5d superconformal field theories (SCFTs), for instance in \cite{Jefferson:2017ahm,Bhardwaj:2019jtr, Hayashi:2018lyv}. On the supergravity front, type IIB solutions dual to brane webs have been put to test by comparison with conformal field theory (CFT) obsevables in \cite{Legramandi:2021uds, Bergman:2018hin} among others. 

A convenient way to realise 5d SCFTs is in terms of 5-brane webs in type IIB string theory \cite{Aharony:1997bh}. The web description is particularly useful for studying the moduli space of vacua. 5d $\mathcal{N}=1$ theories admit a Coulomb branch (CB) which is real and will not play a significant role in our discussion. The Higgs branch (HB) on the other hand is HyperKahler and has been subject to many recent investigations \cite{Akhond:2020vhc, Akhond:2021knl,vanBeest:2020civ,vanBeest:2020kou,Closset:2020scj,Eckhard:2020jyr, Bourget:2019rtl, Bourget:2020gzi, Cabrera:2018jxt, Ferlito:2017xdq, Closset:2020afy, Giacomelli:2020ryy, Carta:2021whq}. The core idea is to encode the Higgs branch chiral ring of a given 5d theory in terms of its magnetic quiver. The magnetic quiver is a 3d $\mathcal{N}=4$ quiver gauge theory whose Coulomb branch is isomorphic to the Higgs branch of the parent 5d theory of interest. The programme's success is due to a flurry of recent results \cite{Cremonesi:2013lqa,Cremonesi:2014xha, Cremonesi:2014kwa, Cremonesi:2014vla} in computing the Coulomb branch Hilbert series of 3d $\mathcal{N}=4$ theories.   

In this paper we initiate a study of the Higgs branch of 5d SCFTs engineered using brane webs with O7$^+$-planes. The relevance of this study is the following. The usual brane web description of 5d SCFTs \cite{Aharony:1997bh} can be enriched with the inclusion of orientifold planes. There are two types of orientifold planes whose inclusion in the brane web leads to consistent 5d SCFTs, namely O5-planes \cite{Zafrir:2015ftn,Zafrir:2016jpu} and O7-planes \cite{Bergman:2015dpa}. Magnetic quivers for 5-brane webs with O5-planes have already been explored in \cite{Akhond:2020vhc, Akhond:2021knl,Bourget:2020gzi}, but an analogous study of brane webs with O7-planes is so far missing from the literature. There are two variants of the O7-plane, namely the O7$^+$ and O7$^-$. The latter of these is non-perturbatively resolved into a pair of $(p,q)$ 7-branes at strong string coupling \cite{Sen:1996vd, Banks:1996nj}, leading therefore to an ordinary brane web. On the other hand the O7$^+$ plane is an exact configuration and so will give rise to novel magnetic quivers.

The strategy for proposing the magnetic quivers is the following. We focus on SO($N$) gauge theories with $N_\textbf{v}$ hypermultiplets in the vector representation. This is a natural set of theories to study, as they admit a realization with a brane-web involving an O7$^+$ plane, and also another realization with a brane-web involving O5$^-$/$\widetilde{\text{O5}}^-$-planes. Since the magnetic quivers for the latter can be obtained using the techniques of \cite{Akhond:2020vhc, Bourget:2020gzi} they serve as a consistency check for our proposal for the magnetic quivers obtained from the web with O7$^+$-plane. 

As a main result, we find that the magnetic quivers for the webs with O7$^+$-planes are always framed non-simply-laced quivers. The derivation of the magnetic quivers was performed with some educated guesswork loosely motivated by intuition stemming from S-duality of type IIB, and some existing results on non-simply laced quivers and brane systems \cite{Cremonesi:2014xha}. 
What gives these conjectural magnetic quivers a firm basis is the agreement of their CB Hilbert series with that computed from the magnetic quivers for the same theory, derived from an O5-plane construction. 

The main result of our study is summarised in table \ref{tab:MQs of SO(N) theory}. Here we illustrate the non-simply-laced quivers which are claimed to be the magnetic quivers for the UV fixed point limit of 5d SO($N$) gauge theory with $N_\textbf{v}$ vectors. One advantage of these magnetic quivers to the orthosymplectic (OSp) quivers obtained from the web with O5-planes is that they are true for even or odd $N$.

\begin{table}[!htb]
    \centering
    \begin{tabular}{|c|c|c|}\hline
         &MQ  &Symmetry\\\hline
  $N_\textbf{v}=N-3$& $\begin{array}{c}
       \begin{scriptsize}
    \begin{tikzpicture}
    \node[label=below:{1}][u](1){};
    \node[label=below:{2}][u](2)[right of=1]{};
    \node (dots)[right of=2]{$\cdots$};
    \node[label=below:{$N_\textbf{v}$}][u](2N-5)[right of=dots]{};
    \node[label=below:{$N_\textbf{v}+1$}][u](2N-4)[right of=2N-5]{};
    \node[uf][right of=2N-4]{};
    \node[label=below:{$2$}][u](2')[right of=2N-4]{};
    \node[label=above:{2}][uf](uf)[above of=2']{};
    \draw(2')--(uf);
    \draw(1)--(2);
    \draw(2)--(dots);
    \draw(dots)--(2N-5);
    \draw[ double distance=1.5pt,<-](2N-5)--(2N-4);
    \draw(2N-4)--(2');
    \end{tikzpicture}
    \end{scriptsize}\end{array}$&$\begin{array}{c}
         \mathfrak{usp}(2N_\textbf{v}+2)
    \end{array}$\\\hline  
    $N_\textbf{v}=N-4$&$\begin{array}{c}
         \begin{scriptsize}
    \begin{tikzpicture}
    \node[label=below:{1}][u](1){};
    \node[label=below:{2}][u](2)[right of=1]{};
    \node (dots)[right of=2]{$\cdots$};
    \node[label=below:{$N_\textbf{v}-1$}][u](2N-5)[right of=dots]{};
    \node[label=below:{$N_\textbf{v}$}][u](2N-4)[right of=2N-5]{};
    \node[uf][right of=2N-4]{};
    \node[label=below:{$2$}][u](2')[right of=2N-4]{};
    \node[label=below:{$1$}][u](1')[right of=2']{};
    \node[label=above:{2}][uf](uf)[above of=2']{};
    \draw(uf)--(2');
    \draw(1)--(2);
    \draw(2)--(dots);
    \draw(dots)--(2N-5);
    \draw[ double distance=1.5pt,<-](2N-5)--(2N-4);
    \draw(2N-4)--(2');
    \draw[double distance=1.5pt,->](2')--(1');
    \end{tikzpicture}
    \end{scriptsize}
    \end{array}$&$\begin{array}{c}
         \mathfrak{usp}(2N_\textbf{v})\oplus\mathfrak{usp}(2)
    \end{array}$\\\hline
   $\begin{array}{c}0\leq N_\textbf{v} \leq N-5\\ N_\textbf{v}\in2\mathbb{Z}+1\end{array}$ &$\begin{array}{c}
             \begin{scriptsize}
    \begin{tikzpicture}
    \node[label=below:{1}][u](1){};
    \node[label=below:{2}][u](2)[right of=1]{};
    \node (dots)[right of=2]{$\cdots$};
    \node[label=below:{$N_\textbf{v}-1$}][u](2N-5)[right of=dots]{};
    \node[label=below:{$N_\textbf{v}$}][u](2N-4)[right of=2N-5]{};
    \node[label=below:{$1$}][u](1')[right of=2N-4]{};
    \node[label=above:{1}][uf](uf)[above of=2N-4]{};
    \node[label=above right:{$3$}][uf](uf')[above of=1']{};
    \node[label=above:{$\left\lceil\frac{N-N_\textbf{v}-6}{2}\right\rceil$}][uf](uf'')[right of=1']{};
    \path [draw,snake it](1')--(uf'');
    \draw(1')--(uf');
    \draw(1)--(2);
    \draw(2)--(dots);
    \draw(dots)--(2N-5);
    \draw[ double distance=1.5pt,<-](2N-5)--(2N-4);
    \draw(2N-4)--(1');
    \draw(2N-4)--(uf);
    \end{tikzpicture}
    \end{scriptsize}
    \end{array}$&$\begin{array}{c}
         \mathfrak{usp}(2N_\textbf{v})\oplus\mathfrak{u}(1)
    \end{array}$\\\hline 
 $\begin{array}{c}0\leq N_\textbf{v} \leq N-5\\ N_\textbf{v}\in2\mathbb{Z}\end{array}$ &$\begin{array}{c}
             \begin{scriptsize}
    \begin{tikzpicture}
    \node[label=below:{1}][u](1){};
    \node[label=below:{2}][u](2)[right of=1]{};
    \node (dots)[right of=2]{$\cdots$};
    \node[label=below:{$N_\textbf{v}-1$}][u](2N-5)[right of=dots]{};
    \node[label=below:{$N_\textbf{v}$}][u](2N-4)[right of=2N-5]{};
    \node[label=below:{$1$}][u](1')[right of=2N-4]{};
    \node[label=above:{1}][uf](uf)[above of=2N-4]{};
    \node[label=above right:{$4-\left\lceil\frac{N}{2}\right\rceil+\left\lfloor\frac{N}{2}\right\rfloor$}][uf](uf')[above of=1']{};
    \node[label=above:{$\left\lceil\frac{N-N_\textbf{v}-6}{2}\right\rceil$}][uf](uf'')[right of=1']{};
    \path [draw,snake it](1')--(uf'');
    \draw(1')--(uf');
    \draw(1)--(2);
    \draw(2)--(dots);
    \draw(dots)--(2N-5);
    \draw[ double distance=1.5pt,<-](2N-5)--(2N-4);
    \draw(2N-4)--(1');
    \draw(2N-4)--(uf);
    \end{tikzpicture}
    \end{scriptsize}
    \end{array}$&$\begin{array}{c}
         \mathfrak{usp}(2N_\textbf{v})\oplus\mathfrak{u}(1)
    \end{array}$\\\hline 
    \end{tabular}
    \caption{Magnetic quivers for the infinite gauge coupling limit of 5d SO$(N)$ gauge theory with $N_\textbf{v}$ hypermultiplets transforming in the vector representation of SO$(N)$. The reader is refered to section \ref{secHS} for the notation we use in the quivers.}
    \label{tab:MQs of SO(N) theory}
\end{table}

The rest of the paper is organised as follows. In section \ref{secHS} we review known results about various aspects of Coulomb branch Hilbert series, and we discuss in particular the peculiar aspects arising in the computation of such HS when the quiver is non-simply laced. Section \ref{sec2MQ} contains the main results of our work. Here we explain how to read off the Higgs branch directions from brane webs with O7$^+$-plane and obtain the magnetic quivers. We then proceed to compute the Hilbert series of the magnetic quivers obtained from brane webs with O7$^+$-plane and match the result with magnetic quivers obtained for the same theory but constructed using an O5-plane. Section \ref{discussion} contains our conclusions and hints for further studies. 
%\bigskip
%--------- Section 2 --------------
\section{Coulomb branch Hilbert series for non-simply laced quivers}\label{secHS}

In this section we recall some facts about the Hilbert series for the Coulomb branch of a $3d$ $\mathcal{N}=4$ gauge theory. After some general remarks, we will discuss peculiar features that arise when the quiver is non-simply laced. We refer the reader to \cite{Cremonesi:2013lqa} for further discussion.

A Coulomb branch Hilbert series counts the number of \emph{dressed} monopole operators, graded by their conformal dimension. Let $G=\prod_{i=1}^n G_i$ be the gauge group of the theory, which we assume for the moment to be described by a simply-laced quiver. A given \emph{bare} monopole operator has magnetic charges $(\vec{m_1},\ldots \vec{m_n})$ taking values in the weight lattice $\Gamma(\hat{G})$ of the GNO dual of the gauge group $\hat{G}$. The conformal dimension of a bare monopole operator of magnetic charges $(\vec{m_1},\ldots \vec{m_n})$ is given by the monopole dimension formula

\begin{equation}
\label{monopoledimension}
\Delta(m)=-\sum_{\alpha\in\Delta_+}\left|\alpha(m)\right|+\dfrac{1}{2}\sum_{i=1}^n\sum_{\rho_i\in\mathcal{R}_i}\left|\rho_i(m)\right| .
\end{equation}
The first sum in (\ref{monopoledimension}) is over the positive roots $\alpha\in\Delta_+$ of the Lie algebra of the gauge group, and correspond to the contributions of the vector multiplets to the monopole dimension. The second sum is over the weights $\rho_i\in\mathcal{R}_i$ of the representations of the gauge Lie algebra in which the hypermultiplets transform. Following the conventions of \cite{Gaiotto:2008ak} we call a $3d$ $\mathcal{N}=4$ theory \emph{good}, if all the monopole operators have dimension $\Delta>\frac{1}{2}$, which is the unitarity bound for scalar operators in a 3d CFT. If all monopole operators have dimension $\Delta\geq\frac{1}{2}$ and in particular some of them saturate the unitarity bound, we call the theory \emph{ugly}. If some monopole operator has dimension $\Delta<\frac{1}{2}$ we call the theory \emph{bad}.

Under the hypothesis that the theory is not bad, the Coulomb Branch Hilbert series reads

\begin{equation}
    HS_C(t)=\sum_{\vec{m_1}}\sum_{\vec{m_2}}\cdots\sum_{\vec{m_n}}t^{\Delta(\vec{m_1},\ldots,\vec{m_n})}\prod_{i=1}^nP_{G_i}(t,\vec{m_i})
    \label{eq:HSgeneric}
\end{equation}
where the sums are performed over the Weyl chambers of the weight lattice $\Gamma^*_{\hat{G}}$ of the GNO dual of the gauge group, and $P_{G_i}(t,\vec{m_i})$ is a dressing factor to take into account the fact that bare monopole operator can be dressed by the scalars in the adjoint of the vectormultiplets of the subgroup $H\subset G$ left unbroken by the monopole. For an explicit discussion of the dressing factor, we defer the reader to the appendix of \cite{Cremonesi:2013lqa}.

When the quiver consists only of unitary nodes, without the addition of flavors, there is typically an overall U$(1)$ which is decoupled. At the level of the HS computation, the decoupling of the overall U$(1)$ can be implemented in two ways. One can either treat one U$(N)$ gauge node as SU$(N)$ since the beginning, or one can treat it as U$(N)$ when computing the monopole dimension formula, and then set to zero one of the magnetic charges of such U$(N)$ at the moment of computing the Hilbert Series. Furthermore, the choice of the specific node at which the overall U$(1)$ decouples is immaterial. We will see that both these features cease to be true if we relax the original hypothesis that the quiver is simply-laced.

Let us consider now a case in which the quiver itself is unitary and non-simply laced. With this we mean that two nodes can be connected by $n$ oriented lines. In the usual non-simply laced case, a straight line between two gauge nodes is interpreted as a bifundamental hypermultiplet. In the non-simply laced case, there is up to date no clear interpretation of the oriented multiple-line, at the field theory level. Despite this, it was proposed in \cite{Cremonesi:2014xha} that one can still compute the Coulomb Branch Hilbert series of such a quiver. First of all, the matter contribution to the dimension formula must be modified as follows\footnote{In case of multiplicity $n$, we replace the number $2$ in $|2m_i-n_j|$ with the number $n$.}:

\begin{equation}
    \Delta_{\text{hyp}}=\left\{\begin{array}{cc}
         \frac{1}{2}\sum_{i=1}^{M}\sum_{j=1}^{N}\lvert2m_i-n_j \rvert & \begin{scriptsize}\begin{tikzpicture}
         \node[label=above:{$M$}][u](N1){};
         \node[label=above:{$N$}][u](N2)[right of=N1]{};
         \draw[ double distance=1.5pt,->](N1)--(N2);
         \end{tikzpicture}  \end{scriptsize} \\
         \frac{1}{2}\sum_{i=1}^{M}\sum_{j=1}^{N} \lvert m_i-n_j \rvert & \begin{scriptsize}
         \begin{tikzpicture}
          \node[label=above:{$M$}][u](N1){};
         \node[label=above:{$N$}][u](N2)[right of=N1]{};
         \draw(N1)--(N2);
         \end{tikzpicture}
         \end{scriptsize}
    \end{array}\right. .
\end{equation}

Secondly, it turns out that for a unitary non-simply laced quiver with no flavor nodes there is still an overall U$(1)$ that needs to be decoupled. However, now the choice of where to decouple the overall U$(1)$ is crucial. Decoupling it at different nodes will result in different CB Hilbert series. Therefore one denotes with the squircle $\begin{array}{c}
     \begin{tikzpicture}
      \node[uf]{};
      \node[u]{};
     \end{tikzpicture}
\end{array}$ the location at which the overall U$(1)$ has to be decoupled. See \cite{Hanany:2020jzl} for more details on this point.

Not just the node at which the decoupling is done, but also \emph{how} this procedure is done is important. In the simply-laced case we could have decoupled the overall U$(1)$ both by treating a U$(N)$ node as an SU$(N)$ when writing the monopole dimension formula, or at a later stage. In this second option, one writes the monopole dimension formula as if all nodes are unitary, and then simply does not sum over one of the magnetic charges when computing (\ref{eq:HSgeneric}), namely, one sets $m_i=0$  for a given fugacity $m_i$. We argue that the two prescriptions are not in general equivalent, and in particular when the quiver is non-simply laced, the second option is the correct one.

The reason for this is very evident working in the fugacity base of U$(1)\times SU(N)$. For simplicity, let us consider $N=2$. Let $q$ be the fugacity associated to U$(1)$ and $p$ that associated to SU$(2)$. The change of base that connects these fugacities to the usual ones is \cite{Hanany:2016ezz}:
\begin{equation}
    \begin{cases}
    m_1=p+q\\
    m_2=p-q
    \end{cases} .
\end{equation}
From this we see that setting to zero $m_1$ means forcing $p=q$. This means decoupling a U$(1)_{\mbox{diag}}$ which is \emph{not} the U$(1)$ factor in the product U$(1)\times$ SU$(2)$, but rather it is a diagonal U$(1)_{\mbox{diag}}$  between the U$(1)$ factor of U$(1)\times$ SU$(2)$ and the Cartan U$(1)_{\mbox{car}}$ of SU$(2)$. On the other hand, treating the U$(2)$ node as SU$(2)$ from the beginning would correspond to set to zero $p$ when writing the monopole dimension formula. The two operations are clearly different. In the simply laced case the end result of the HS computation will not depend on which option is chosen, however this is just an accident. In the simply-laced case the result will depend on this choice, and throughout this paper we find consistent results if and only if the diagonal U$(1)$ is the one to be ungauged.

Finally, in this paper, we will also use hypermultiplets transforming under the charge 2 representation of U(1) gauge nodes. In the quiver, we denote $F$ of such hypermultiplets by a wiggly line. In particular we write
\begin{equation}
    \begin{array}{c}
         \begin{scriptsize}
             \begin{tikzpicture}
                    \node[label=below:{1}][u](u1){};
                    \node[label=below:{$F$}][uf](uf)[right of=u1]{};
                    \path [draw,snake it](u1)--(uf);
             \end{tikzpicture}
         \end{scriptsize}
    \end{array}\;.
\end{equation}
%to denote $F$ hypermultiplets transforming under the charge 2 representation of U(1).

%--------- Section 3 --------------
\section{Magnetic quivers from O7-planes}\label{sec2MQ}
In this section we discuss how one can read off the Higgs branch directions from brane webs with O7$^+$-planes. We then combine this knowledge with known facts about the Higgs branch of SO($N$) gauge theories at infinite coupling, as well as existing results on non-simply laced quivers and brane constructions to fix the structure of the corresponding magnetic quivers. We then provide consistency checks by computing the HS of the proposed magnetic quivers and matching with the corresponding computation on OSp magnetic quivers derived from brane webs with O5-planes.
\subsection{Pure SO(N) theory}
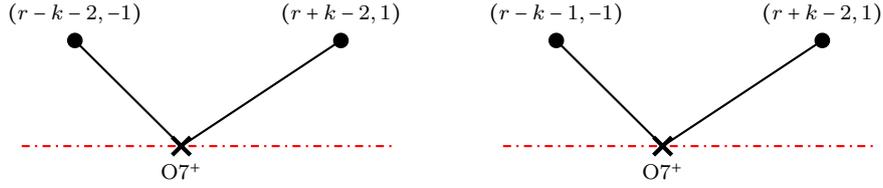
\begin{figure}[!htb]
    \centering
    \begin{scriptsize}
    \begin{tikzpicture}[scale=.7]
    \draw[thick,dash dot,red](-3,0)--(4,0);
    \node[label=below:{O7$^+$}][cross]at (0,0){};
    \draw[thick](0,0)--(-2,2);
    \node[label=above:{$(r-k-2,-1)$}][7brane]at (-2,2){};
    \draw[thick](0,0)--(3,2);
    \node[label=above:{$(r+k-2,1)$}][7brane]at (3,2){};
    \end{tikzpicture}
    \end{scriptsize}
    \hspace{.5cm}
    \begin{scriptsize}
    \begin{tikzpicture}[scale=.7]
    \draw[thick,dash dot,red](-3,0)--(4,0);
    \node[label=below:{O7$^+$}][cross]at (0,0){};
    \draw[thick](0,0)--(-2,2);
    \node[label=above:{$(r-k-1,-1)$}][7brane]at (-2,2){};
    \draw[thick](0,0)--(3,2);
    \node[label=above:{$(r+k-2,1)$}][7brane]at (3,2){};
    \end{tikzpicture}
    \end{scriptsize}
    \caption{Brane webs for pure SO($N$) gauge theory at the infinite gauge coupling limit for the case where $N=2r$ (left) and $N=2r+1$ (right).}
    \label{fig:pure SO(N)}
\end{figure}
The brane web for pure SO($N$) gauge theory at the infinite gauge coupling limit using an O7-plane was first proposed in \cite{Bergman:2015dpa} and is shown in figure \ref{fig:pure SO(N)}. The integer $k$ labels the choice of $SL(2,\mathbb{Z})$ frame, and in particular under the generator $T$, each of the web diagrams in figure \ref{fig:pure SO(N)} is mapped to itself with $k\rightarrow k+1$. Note that the brane web looks slightly different depending on whether the gauge algebra is of B-type or D-type. We would first like to understand how to read off the Higgs branch directions from these web diagrams. For the SO($2r$) theory in figure \ref{fig:pure SO(N)}, one can separate the $(r-k-2,-1)$ and the $(r+k-2,1)$ 5-branes along the Higgs branch directions giving rise to a Higgs branch of unit quaternionic dimension. That this is a one-dimensional space rather than a two dimensional one is due to the fact that the moduli correspond to the relative positions of the two independent subwebs. Equivalently one has the freedom to fix the position of one of the subwebs to the origin of the transverse space. This is unlike the situation one would encounter when dealing with brane webs with O5-planes due to the fact that the O5-plane provides a refrence point, and the number of available directions is equal to the number of independent subwebs. But since an O7-plane spans all the directions transverse to the plane in which the web is drawn, i.e. it is not localised to any point along the Higgs branch directions it cannot serve as a refrence point and the total available Higgs branch directions are one less than the number of independent subwebs. For the SO($2r+1$) theory a similar statement holds for the $(r-k-1,-1)$ and the $(r+k-2,1)$ 5-branes. The naive prescription for obtaining the magnetic quiver, had the O7$^+$-plane not been present, would be \cite{Cabrera:2018jxt} to assign a U(1) gauge node to each subweb and connect them together by as many hypermultiplets as the Schwinger product, or bare stable intersection\footnote{We refer the reader to \cite{Cabrera:2018jxt} for the precise definition} of the $(p,q)$ charges of the independent subwebs in figure \ref{fig:pure SO(N)}, with the knowledge that an overall U(1) acts trivially, corresponding to the freedom to fix the position of one independent subweb to the origin. On the other hand, it is known \cite{Cremonesi:2015lsa} that the Higgs branch of the infinite gauge coupling limit of a 5d $\mathcal{N}=1$ super Yang-Mills theory with gauge group $G$ should be given by the orbifold $\mathbb{C}^2/\mathbb{Z}_{h_G^\vee}$, where $h_G^\vee$ denotes the dual Coxeter number of the group $G$. Moreover, the Coulomb branch of 3d $\mathcal{N}=4$ QED with $n_1$ charge one hypermultiplets and $n_2$ charge 2 hypermultiplets is $\mathbb{C}^2/\mathbb{Z}_{n_1+2n_2}$. If we further require the Higgs branch dimension of the magnetic quiver to be equal to the rank of the 5d theory, then the magnetic quiver is uniquely determined to be 
\begin{equation}\label{MQ for pure SO(N)}
    \mathcal{H_\infty}\left(\text{SO}(N)\right)=\mathbb{C}^2/\mathbb{Z}_{h_{\text{SO}(N)}^\vee}=\mathcal{C}^{3\text{d}}\left(\begin{array}{c}
         \begin{scriptsize}
         \begin{tikzpicture}
         \node[label=below:{1}][u](u1){};
         \node[label=right:{$\left\lceil\frac{N-6}{2}\right\rceil$}][uf](uf)[right of=u1]{};
         \node[label=above:{$4-\left\lceil\frac{N}{2}\right\rceil+\left\lfloor\frac{N}{2}\right\rfloor$}][uf](uf')[above of=u1]{};
         \draw(u1)--(uf');
         \path [draw,snake it](u1)--(uf);
         \end{tikzpicture}
         \end{scriptsize}
    \end{array}\right)\;,
\end{equation}
where $\left\lfloor x \right\rfloor$, $\left\lceil x \right\rceil$ denote the floor and ceiling function respectively, and are defined by
\begin{equation}
    \left\lfloor x \right\rfloor=\max\left\{n\in\mathbb{Z}\;|\;n\leq x\right\}\;,\qquad
    \left\lceil x \right\rceil=\min\left\{n\in\mathbb{Z}\;|\;n\geq x\right\}\;.
\end{equation}
Our task now is to recover \eqref{MQ for pure SO(N)} from the data in the brane webs of figure \ref{fig:pure SO(N)}. Let us first take the Schwinger product of the $(p,q)$ charges of the two independent subwebs in each of the webs in figure \ref{fig:pure SO(N)} to find
\begin{equation}\label{bare SI}
\begin{aligned}
    \left| \det\begin{pmatrix} (r-k-2) & -1\\
    (r+k-2) & 1
    \end{pmatrix} \right|&=2r-4\\
    \left| \det\begin{pmatrix} (r-k-1) & -1\\
    (r+k-2) & 1
    \end{pmatrix} \right|&=2r-3
\end{aligned}
\end{equation}
Upon comparison with \eqref{MQ for pure SO(N)} we propose that the number of charge 2 hypermultiplets be given by the formula
\begin{equation}\label{stable intersection}
n_\text{charge 2}=\left\lceil\frac{\text{SI}_0-2}{2}\right\rceil \;,
\end{equation}
where SI$_0$ refers to the bare stable intersection number, i.e. those computed in \eqref{bare SI}. The number of charge one hypermultiplets, is fixed to be 4 in the case of $N=2r$ and 3 in the case when $N=2r+1$. We do not currently have a good microscopic understanding of these hypermultiplets, but this should not stop us from proposing the magnetic quivers. At this point the reader might question our assumption that only charge 1 or charge 2 hypers could appear in the magnetic quiver. However, in the next section we provide examples where an OSp dual is known and comparison of the HS of the non-simply-laced and OSp quivers leads to a justification of this assumption.
\subsection{\texorpdfstring{SO$(N)$ with $N_\textbf{v}\leq N-5$}{TEXT}}
\begin{figure}[!htb]
    \centering
    \begin{scriptsize}
\begin{tikzpicture}[scale=.7]
    \draw[thick,dash dot,red](-3,0)--(4,0);
    \node[label=below:{O7$^+$}][cross]at (0,0){};
    \draw[thick](0,0)--(-2,2);
    \node[label=above:{$(r-k-2,-1)$}][7brane]at (-2,2){};
    \draw[thick](0,0)--(3,2);
    \node[label=above:{$(r+k-2,1)$}][7brane]at (3,2){};
    \node[7brane]at (0.25,1){};
    \node[7brane]at (-.25,1){};
    \node[7brane]at (.75,1){};
    \draw [thick,decorate,decoration={brace,amplitude=6pt},xshift=0pt,yshift=10pt]
(-.5,1) -- (1,1)node [black,midway,xshift=0pt,yshift=12pt] {
\footnotesize $N_\textbf{v}$ D7};
    \end{tikzpicture}
    \end{scriptsize}
    \hspace{.5cm}
    \begin{scriptsize}
    \begin{tikzpicture}[scale=.7]
    \draw[thick,dash dot,red](-3,0)--(4,0);
    \node[label=below:{O7$^+$}][cross]at (0,0){};
    \draw[thick](0,0)--(-2,2);
    \node[label=above:{$(r-k-1,-1)$}][7brane]at (-2,2){};
    \draw[thick](0,0)--(3,2);
    \node[label=above:{$(r+k-2,1)$}][7brane]at (3,2){};
    \node[7brane]at (0.25,1){};
    \node[7brane]at (-.25,1){};
    \node[7brane]at (.75,1){};
    \draw [thick,decorate,decoration={brace,amplitude=6pt},xshift=0pt,yshift=10pt]
(-.5,1) -- (1,1)node [black,midway,xshift=0pt,yshift=12pt] {
\footnotesize $N_\textbf{v}$ D7};
    \end{tikzpicture}
    \end{scriptsize}
    \caption{Brane webs for pure SO($N$) gauge theory with $N_\textbf{v}$ hypermultiplets in the vector representation at the infinite gauge coupling limit for the case where $N=2r$ (left) and $N=2r+1$ (right).}
    \label{fig:brane web with flavour D7s}
\end{figure}
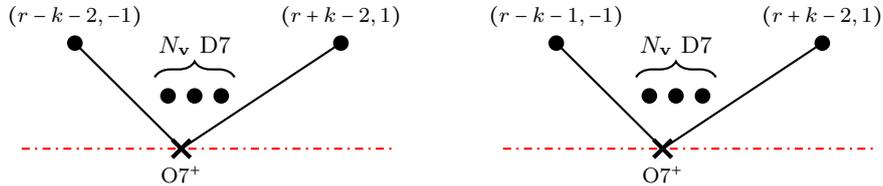
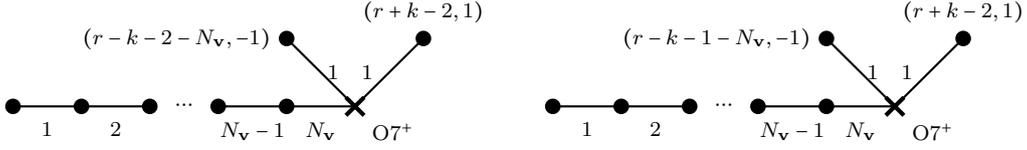
\begin{figure}
    \centering
    \begin{scriptsize}
    \begin{tikzpicture}[scale=.9]
    \draw[thick](-1,-1)--(0,-2);
    \draw[thick](0,-2)--(1,-1);
    \node[label=above:{$(r+k-2,1)$}][7brane] at (1,-1){};
    \node[label=left:{$(r-k-2-N_\textbf{v},-1)$}][7brane] at (-1,-1){};
    \node[label=below right:{O7$^+$}][cross] at (0,-2){};
    \draw[thick](0,-2)--(-2,-2);
    \draw[thick](-5,-2)--(-3,-2);
    \node at (-2.5,-2){$\cdots$};
    \node[7brane] at (-1,-2){};
    \node[7brane] at (-2,-2){};
    \node[7brane] at (-3,-2){};
    \node[7brane] at (-4,-2){};
    \node[7brane] at (-5,-2){};
    \node [label=below:{1}] at (-4.5,-2){};
    \node [label=below:{2}] at (-3.5,-2){};
    \node [label=below:{$N_\textbf{v}-1$}] at (-1.5,-2){};
    \node [label=below:{$N_\textbf{v}$}] at (-.5,-2){};
    \node[label=left:{1}] at (0,-1.5){};
    \node[label=left:{1}] at (0.5,-1.5){};
    \end{tikzpicture}
    \end{scriptsize}
    \hspace{.25cm}
    \begin{scriptsize}
    \begin{tikzpicture}[scale=.9]
    \draw[thick](-1,-1)--(0,-2);
    \draw[thick](0,-2)--(1,-1);
    \node[label=above:{$(r+k-2,1)$}][7brane] at (1,-1){};
    \node[label=left:{$(r-k-1-N_\textbf{v},-1)$}][7brane] at (-1,-1){};
    \node[label=below right:{O7$^+$}][cross] at (0,-2){};
    \draw[thick](0,-2)--(-2,-2);
    \draw[thick](-5,-2)--(-3,-2);
    \node at (-2.5,-2){$\cdots$};
    \node[7brane] at (-1,-2){};
    \node[7brane] at (-2,-2){};
    \node[7brane] at (-3,-2){};
    \node[7brane] at (-4,-2){};
    \node[7brane] at (-5,-2){};
    \node [label=below:{1}] at (-4.5,-2){};
    \node [label=below:{2}] at (-3.5,-2){};
    \node [label=below:{$N_\textbf{v}-1$}] at (-1.5,-2){};
    \node [label=below:{$N_\textbf{v}$}] at (-.5,-2){};
    \node[label=left:{1}] at (0,-1.5){};
    \node[label=left:{1}] at (0.5,-1.5){};
    \end{tikzpicture}
    \end{scriptsize}
    \caption{Brane webs for SO$(N)$+$N_\textbf{v}$ hypermultiplets in the vector representation at infinite gauge coupling after taking all mass parameters to zero for $N=2r$ (left) and $N=2r+1$ (right). From here onwards we will not indicate the monodromy cut associated with the 7-brane explicitly for ease of presentation.}
    \label{fig:SO(N)+Nv<5 web}
\end{figure}
Having gained some intuition about the Higgs branch of pure SO($N$) theory at infinite coupling, we can now proceed with more involved cases. We can add matter in the vector representation of SO$(N)$ by adding D7 branes to the brane configuration as in figure \ref{fig:brane web with flavour D7s}. To read off the magnetic quiver, we first perform a few Hanany-Witten moves \cite{Hanany:1997gh} by moving all flavour D7s, say, to the left of the $(r-k-2,-1)$ (resp. $(r-k-1,-1)$) 5-brane in figure \ref{fig:brane web with flavour D7s}. This creates $N_\textbf{v}$ D5 branes and further changes the charge of the $(r-k-2,-1)$ (resp. $(r-k-1,-1)$) 7-brane to $(r-k-2-N_\textbf{v},-1)$ (resp. $(r-k-1-N_\textbf{v},-1)$) for $N=2r$ (resp. $N=2r+1$). As a final step we set all the mass parameters in the web to zero and separate the D7 branes along the horizontal direction in the web diagram. The end result of this process is depicted in figure \ref{fig:SO(N)+Nv<5 web}. We immediately identify the independent subwebs to be the D5 branes extended between the adjacent pairs of D7 branes, D5 branes extended between the O7$^+$-plane and the D7 closest to it in addition to the $(r+k-2,1)$ 5-brane and the $(r-k-2-N_\textbf{v},-1)$ (resp. $(r-k-1-N_\textbf{v},-1)$) for $N=2r$ (resp. $N=2r+1$). Each of the aforementioned subwebs corresponds to a unitary gauge node whose rank is determined by the number of 5-branes in the stack. We need to decouple an overall U(1) to reflect the fact that the Higgs branch directions only care about the relative positions of the aforementioned independent subwebs and we are free to fix the position of one of the subwebs to be at the origin. We will decouple the U(1) node corresponding to the $(r+k-2,1)$ node.  There is a single link connecting any two nodes corresponding to subwebs which end on the same 7-brane from the opposite side, except for those lying on either side of the 7-brane immediately to the left of the O7$^+$-plane in figure \ref{fig:SO(N)+Nv<5 web}, which are connected by a double bond. This is essentially the T-dual of the brane system and magnetic quiver in \cite{Cremonesi:2014xha}. There are charge 1 and charge 2 hypermultiplets attached to nodes corresponding to subwebs that end on the O7$^+$. The number of charge 2 hypers is given by the formula \eqref{stable intersection}, where we take their SI with the subweb which we have frozen to the origin. The number of charge 1 hypers are fixed by requiring the HS to agree with dual OSp quivers, to be mentioned momentarily. The magnetic quiver that we propose in this case is slightly different depending on whether the number $N_\textbf{v}$ of hypermultiplets is even or odd. We propose that for $N_\textbf{v}$ even the magnetic quiver is given by
\begin{equation}\label{MQ N<N-5 Nv even}
    \begin{array}{c}
             \begin{scriptsize}
    \begin{tikzpicture}
    \node[label=below:{1}][u](1){};
    \node[label=below:{2}][u](2)[right of=1]{};
    \node (dots)[right of=2]{$\cdots$};
    \node[label=below:{$N_\textbf{v}-1$}][u](2N-5)[right of=dots]{};
    \node[label=below:{$N_\textbf{v}$}][u](2N-4)[right of=2N-5]{};
    \node[label=below:{$1$}][u](1')[right of=2N-4]{};
    \node[label=above:{1}][uf](uf)[above of=2N-4]{};
    \node[label=above right:{$4-\left\lceil\frac{N}{2}\right\rceil+\left\lfloor\frac{N}{2}\right\rfloor$}][uf](uf')[above of=1']{};
    \node[label=above:{$\left\lceil\frac{N-N_\textbf{v}-6}{2}\right\rceil$}][uf](uf'')[right of=1']{};
    \path [draw,snake it](1')--(uf'');
    \draw(1')--(uf');
    \draw(1)--(2);
    \draw(2)--(dots);
    \draw(dots)--(2N-5);
    \draw[ double distance=1.5pt,<-](2N-5)--(2N-4);
    \draw(2N-4)--(1');
    \draw(2N-4)--(uf);
    \end{tikzpicture}
    \end{scriptsize}
    \end{array}\;;\quad N_\textbf{v}\in 2\mathbb{Z}
\end{equation}
while for odd $N_\textbf{v}$ it should be given by
\begin{equation}\label{MQ Nv<N-5}
\begin{array}{c}
    \begin{scriptsize}
    \begin{tikzpicture}
    \node[label=below:{1}][u](1){};
    \node[label=below:{2}][u](2)[right of=1]{};
    \node (dots)[right of=2]{$\cdots$};
    \node[label=below:{$N_\textbf{v}-1$}][u](2N-5)[right of=dots]{};
    \node[label=below:{$N_\textbf{v}$}][u](2N-4)[right of=2N-5]{};
    \node[label=below:{$1$}][u](1')[right of=2N-4]{};
    \node[label=above:{1}][uf](uf)[above of=2N-4]{};
    \node[label=above right:{$3$}][uf](uf')[above of=1']{};
    \node[label=right:{$\left\lceil\frac{N-N_\textbf{v}-6}{2}\right\rceil$}][uf](uf'')[right of=1']{};
    \path [draw,snake it](1')--(uf'');
    \draw(1')--(uf');
    \draw(1)--(2);
    \draw(2)--(dots);
    \draw(dots)--(2N-5);
    \draw[double distance=1pt,<-](2N-5)--(2N-4);
    \draw(2N-4)--(1');
    \draw(2N-4)--(uf);
    \end{tikzpicture}
    \end{scriptsize}\end{array}\;;\quad N_\textbf{v}\in 2\mathbb{Z}+1\;.
\end{equation}
Note that except for the rightmost U(1) node, all other nodes in these quivers are balanced. This suggests that the Coulomb branch isometry of this quiver is $\mathfrak{usp}(2N_\textbf{v})\oplus \mathfrak{u}(1)$. From the 5d point of view this would suggest no enhancement of the global symmetry at the UV fixed point.
For future comparison, with orthosymplectic quivers, we compute the Coulomb branch Hilbert series for this quiver for $N_\textbf{v}=2,4$ and various values of $N$. For $N_\textbf{v}=2$ the results are as follows
\begin{equation}
\begin{array}{c}
    \begin{tabular}{|c|c|}\hline
         $\begin{array}{c}
              N
         \end{array}$& $\begin{array}{c}
              \HS_{N_\textbf{v}=2}
         \end{array}$ \\\hline
         $\begin{array}{c}
              7
         \end{array}$&$\begin{array}{c}
            1 + 11 t^2 + 60 t^4 + 10 t^5 + 225 t^6 + 80 t^7 + 665 t^8 + 350 t^9 + 
 1694 t^{10} + \mathcal{O}(t^{11})
         \end{array}$\\\hline
         $\begin{array}{c}
              8
         \end{array}$&$\begin{array}{c}
             1 + 11 t^2 + 60 t^4 + 10 t^5 + 225 t^6 + 80 t^7 + 665 t^8 + 350 t^9 + 
 1694 t^{10}+ \mathcal{O}(t^{11})
         \end{array}$\\\hline
         $\begin{array}{c}
              9
         \end{array}$&$\begin{array}{c}
              1 + 11 t^2 + 60 t^4 + 225 t^6 + 10 t^7 + 665 t^8 + 80 t^9 + 1666 t^{10}+ \mathcal{O}(t^{11})
         \end{array}$\\\hline
                  $\begin{array}{c}
              10
         \end{array}$&$\begin{array}{c}
              1 + 11 t^2 + 60 t^4 + 225 t^6 + 10 t^7 + 665 t^8 + 80 t^9 + 1666 t^{10}+ \mathcal{O}(t^{11})
         \end{array}$\\\hline
    \end{tabular}
    \end{array}\;.\label{HS Nv=2 Nv<N-5 non-simply-laced}
\end{equation}

For $N_\textbf{v}=4$ the Hilbert series reads
\begin{equation}
\begin{array}{c}
     \begin{tabular}{|c|c|}\hline
          $\begin{array}{c}
               N
          \end{array}$& $\begin{array}{c}
               \HS_{N_\textbf{v}=4}
          \end{array}$ \\\hline
          $\begin{array}{c}
               9
          \end{array}$&$\begin{array}{c}
               1 + 37 t^2 + 675 t^4 + 8130 t^6+84t^7+73047t^8+\mathcal{O}(t^9)
          \end{array}$\\\hline 
          $\begin{array}{c}
               10
          \end{array}$&$\begin{array}{c}
               1 + 37 t^2 + 675 t^4 + 8130 t^6 + 73131 t^8 +  526815 t^{10} + \mathcal{O}(t^{11})
          \end{array}$\\\hline
          $\begin{array}{c}
               11
          \end{array}$&$\begin{array}{c}
               1 + 37 t^2 + 675 t^4 + 8130 t^6+73047t^8+\mathcal{O}(t^9)
          \end{array}$\\\hline
          $\begin{array}{c}
               12
          \end{array}$&$\begin{array}{c}
               1 + 37 t^2 + 675 t^4 + 8130 t^6+73047t^8+524505 t^{10}+\mathcal{O}(t^{11})
          \end{array}$\\\hline
     \end{tabular}
\end{array}
\end{equation}

\subsubsection{\texorpdfstring{$N=2r$}{TEXT}}
An immediate consistency check to see whether our conjectured magnetic quiver for SO(N) with $N_\textbf{v}\leq N_5$ \eqref{MQ N<N-5 Nv even}, \eqref{MQ Nv<N-5} is to compare its Coulomb branch Hilbert series with the orthosymplectic magnetic quiver for the same theory obtained from an O5-plane construction. Let us focus on $N=2r$ as this is the best understood case. The brane web for SO$(2r)$ gauge theory with $N_\textbf{v}\leq N-5$ takes on a slightly different form depending on whether $N_\textbf{v}$ is odd or even. In the case when $N_\textbf{v}\in 2\mathbb{Z}$ the brane web is given by
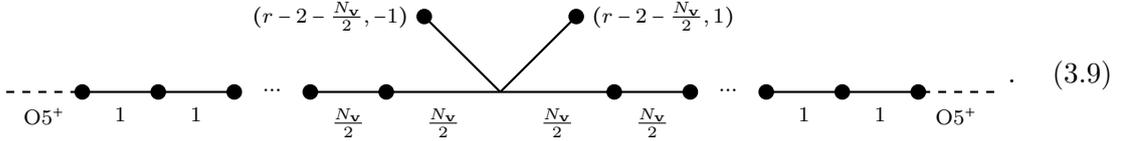
\begin{equation}
    \begin{array}{c}
         \begin{scriptsize}
         \begin{tikzpicture}
         \draw[thick,dashed](0,0)--(1,0);
         \draw[thick](1,0)--(3,0);
         \draw[thick](4,0)--(9,0);
         \draw[thick](12,0)--(10,0);
         \draw[thick](6.5,0)--(7.5,1);
         \draw[thick](6.5,0)--(5.5,1);
         \draw[thick,dashed](13,0)--(12,0);
         \node[7brane]at(1,0){};
         \node[7brane]at(2,0){};
         \node[7brane]at(3,0){};
         \node[7brane]at(4,0){};
         \node[7brane]at(5,0){};
         \node[7brane]at(9,0){};
         \node[7brane]at(8,0){};
         \node[7brane]at(10,0){};
         \node[7brane]at(11,0){};
         \node[7brane]at(12,0){};
         \node[label=left:{$(r-2-\frac{N_\textbf{v}}{2},-1)$}][7brane]at(5.5,1){};
         \node[label=right:{$(r-2-\frac{N_\textbf{v}}{2},1)$}][7brane]at(7.5,1){};
         \node at (3.5,0){$\cdots$};
         \node at (9.5,0){$\cdots$};
         \node[label=below:{O5$^+$}] at (.5,0){};
         \node[label=below:{O5$^+$}] at (12.5,0){};
         \node[label=below:{$1$}] at (1.5,0){};
         \node[label=below:{$1$}] at (2.5,0){};
         \node[label=below:{$\frac{N_\textbf{v}}{2}$}] at (4.5,0){};
         \node[label=below:{$\frac{N_\textbf{v}}{2}$}] at (5.75,0){};
         \node[label=below:{$\frac{N_\textbf{v}}{2}$}] at (7.25,0){};
         \node[label=below:{$\frac{N_\textbf{v}}{2}$}] at (8.5,0){};
         \node[label=below:{$1$}] at (10.5,0){};
         \node[label=below:{$1$}] at (11.5,0){};
         \end{tikzpicture}
         \end{scriptsize}
    \end{array}\;.
\end{equation}
The magnetic quiver that one reads off from this brane web using the methods of \cite{Akhond:2020vhc,Akhond:2021knl} is
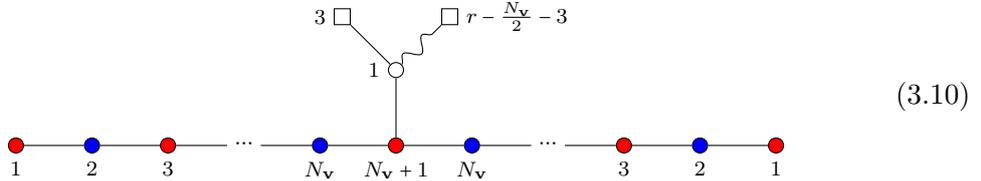
\begin{equation}
    \begin{array}{c}
         \begin{scriptsize}
         \begin{tikzpicture}
         \node[label=below:{1}][so](o1){};
         \node[label=below:{2}][sp](sp1)[right of=o1]{};
         \node[label=below:{3}][so](so3)[right of=sp1]{};
         \node (dots)[right of=so3]{$\cdots$};
         \node[label=below:{$N_\textbf{v}$}][sp](spn-3)[right of=dots]{};
         \node[label=below:{$N_\textbf{v}+1$}][so](so2n-5)[right of=spn-3]{};
         \node[label=below:{$N_\textbf{v}$}][sp](spn-3')[right of=so2n-5]{};
         \node (dots')[right of=spn-3']{$\cdots$};
         \node[label=below:{3}][so](so3')[right of=dots']{};
         \node[label=below:{2}][sp](sp1')[right of=so3']{};
         \node[label=below:{1}][so](o1')[right of=sp1']{};
         \node[label=left:{1}][u](u1)[above of=so2n-5]{};
         \node[label=left:{$3$}][uf](uf)[above left of=u1]{};
         \node[label=right:{$r-\frac{N_\textbf{v}}{2}-3$}][uf](uff)[above right of=u1]{};
         \path [draw,snake it](u1)--(uff);
         \draw(o1)--(sp1);
         \draw(sp1)--(so3);
         \draw(so3)--(dots);
         \draw(dots)--(spn-3);
         \draw(spn-3)--(so2n-5);
         \draw(so2n-5)--(spn-3');
         \draw(spn-3')--(dots');
         \draw(dots')--(so3');
         \draw(so3')--(sp1');
         \draw(sp1')--(o1');
         \draw(so2n-5)--(u1);
         \draw(u1)--(uf);
         \end{tikzpicture}
         \end{scriptsize}
    \end{array}
\end{equation}
Let us compute its Coulomb branch Hilbert series for $N_\textbf{v}=2$. Following the Hall-Littlewood and gluing technique developed in \cite{Cremonesi:2014kwa, Cremonesi:2014vla}, the expression we need to evaluate is the following 
\begin{equation}\label{HS OSp Nv<N-5}
    \sum_{m=0}^{\infty}P_{\text{SU}(2)}(m,t)\HS^2_{T\left[\text{SO}(3)\right]}(m,t)\left(\sum_{-\infty}^{-m}+\sum_{-m+1}^{0}+\sum_{1}^{m}+\sum_{m+1}^{\infty}\right)\frac{t^{\Delta(m,n,N)}}{1-t}\;,
\end{equation}
where $P_{\text{SU}(2)}$ is the dressing factor for the central SO(3) gauge group, the factor $\frac{1}{1-t}$ is the dressing factor for the U(1) gauge group. The Hilbert series for each T[SO(3)] leg in the presence of a background magnetic charge is given by
\begin{equation}
                \HS_{T\left[\text{SO}(3)\right]}(m,t)=\frac{t^{\frac{m}{2}} (1 + m + t - m t)}{(1 - t)^2}\;.
            \end{equation}
            Finally, the conformal dimension, for generic values of $N=2r$ takes the form
\begin{align}
    \Delta(m,n,N)&=-|m|+\frac{1}{2}\left(|n - m| + |n + m| + |n| + (N - 5) |n|\right)\\&=\left\{\begin{array}{cc}
                -m-\frac{n N}{2}+n &-\infty< n\leq -m  \\
                - (N-4)\frac{n}{2}  & m< n\leq 0\\
                 (N-4)\frac{n}{2}  & 0< n \leq m \\
                 (N-2)\frac{n}{2} -m & m< n<\infty
            \end{array} \right.
            \end{align}
            Putting all of this together and evaluating the summations in \eqref{HS OSp Nv<N-5} we obtain the following:

\begin{equation}
\begin{aligned}
    \HS_{N_\textbf{v}=2}(t,N)=\frac{1}{(1 - t^2)^7 (1 - t^{N-2 })^4}\left[1 + 4 t^2 + 4 t^4 + t^6 + 6 t^{N-2} - 16 t^{N+2} - 
 4 t^{N+4} \right.\\- 6 t^{2N-4}
 - 22 t^{2N-2} - 6 t^N + 22 t^{2N} - 
 6 t^{3 N} + 6 t^{2N+2} \\\left.+ 4 t^{3 N-6} + 
 16 t^{3 N-4} + 6 t^{3 N-2} - t^{ 4 N-8} - 
 4 t^{4 N-6} - 
 4 t^{4 N-4} - t^{4 N-2}\right] .
\end{aligned}
\end{equation}

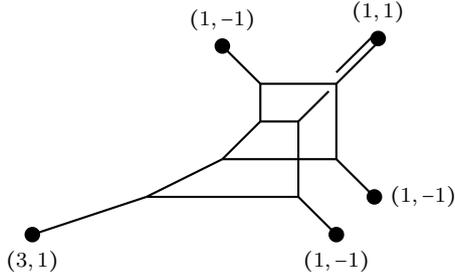
\begin{figure}[!htb]
    \centering
    \begin{scriptsize}
    \begin{tikzpicture}
    \draw[thick] (0,1)--(0,2);
\draw[thick] (0,2)--(-.5,2);
\draw[thick] (0.5,2.5)--(-.5,2.5);
\draw[thick] (-2,1)--(0,1);
\draw[thick] (-.5,2.5)--(-.5,2);
\draw[thick] (-.5,2.5)--(-1,3);
\node[label=above:{$(1,-1)$}][7brane] at (-1,3) {};
\node[label=above:{$(1,1)$}][7brane] at (1.05,3.1) {};
\draw[thick] (-.5,2)--(-1,1.5);
\draw[thick] (-1,1.5)--(-2,1);
\draw[thick] (-3.5,.5)--(-2,1);
\node[label=below:{$(3,1)$}][7brane] at (-3.5,.5) {};
\draw[thick] (0,2)--(.4,2.4);
\draw[thick] (.5,2.5)--(1.1,3.1);
\draw[thick] (.5,2.65)--(1.,3.15);

\draw[thick] (0.5,1.5)--(0.5,2.5);
\draw[thick] (-1,1.5)--(.5,1.5);
\draw[thick] (1,1)--(.5,1.5);
\node[label=right:{$(1,-1)$}][7brane] at (1,1) {};
\node[label=below:{$(1,-1)$}][7brane] at (.5,.5) {};
\draw[thick] (.5,.5)--(0,1);
    \end{tikzpicture}
    \end{scriptsize}
    \caption{Brane web for SU(4)$_0$+1\textbf{AS}.}
    \label{fig:SU(4)_0 + 1AS}
\end{figure}

A similar computation for the $N_\textbf{v}=4$ case leads to 
\begin{equation}
    \begin{array}{c}
         \HS\rvert_{N=10}=1 + 37 t^2 + 675 t^4 + 8130 t^6 + 73131 t^8 + 526815 t^{10} + 
 3179939 t^{12} +  \mathcal{O}(t^{13})  \\
 \HS\rvert_{N=12}=1 + 37 t^2 + 675 t^4 + 8130 t^6 + 73047 t^8 + 524505 t^{10} + 
 3147881 t^{12} +  \mathcal{O}(t^{13})\\
 \HS\rvert_{N=14}=1 + 37 t^2 + 675 t^4 + 8130 t^6 + 73047 t^8 + 524421 t^{10} + 
 3145571 t^{12} +  \mathcal{O}(t^{13})
    \end{array}
\end{equation}
The brane web for SO$(2r)$ gauge theory with $N_\textbf{v}\leq N-5$ and $N_\textbf{v}\in 2\mathbb{Z}+1$ is given by
\begin{equation}
    \begin{array}{c}
         \begin{scriptsize}
         \begin{tikzpicture}
         \draw[thick,dashed](0,0)--(1,0);
         \draw[thick](1,0)--(3,0);
         \draw[thick](4,0)--(9,0);
         \draw[thick](12,0)--(10,0);
         \draw[thick](6.5,0)--(7.5,1);
         \draw[thick](6.5,0)--(5.5,1);
         \draw[thick,dashed](13,0)--(12,0);
         \node[7brane]at(1,0){};
         \node[7brane]at(2,0){};
         \node[7brane]at(3,0){};
         \node[7brane]at(4,0){};
         \node[7brane]at(5,0){};
         \node[7brane]at(9,0){};
         \node[7brane]at(8,0){};
         \node[7brane]at(10,0){};
         \node[7brane]at(11,0){};
         \node[7brane]at(12,0){};
         \node[label=left:{$(r-2-\frac{N_\textbf{v}+1}{2},-1)$}][7brane]at(5.5,1){};
         \node[label=right:{$(r-2-\frac{N_\textbf{v}-1}{2},-1)$}][7brane]at(7.5,1){};
         \node at (3.5,0){$\cdots$};
         \node at (9.5,0){$\cdots$};
         \node[label=below:{O5$^+$}] at (.5,0){};
         \node[label=below:{O5$^+$}] at (12.5,0){};
         \node[label=below:{$1$}] at (1.5,0){};
         \node[label=below:{$1$}] at (2.5,0){};
         \node[label=below:{$\frac{N_\textbf{v}-1}{2}$}] at (4.5,0){};
         \node[label=below:{$\frac{N_\textbf{v}-1}{2}$}] at (5.75,0){};
         \node[label=below:{$\frac{N_\textbf{v}+1}{2}$}] at (7.25,0){};
         \node[label=below:{$\frac{N_\textbf{v}+1}{2}$}] at (8.5,0){};
         \node[label=below:{$1$}] at (10.5,0){};
         \node[label=below:{$1$}] at (11.5,0){};
         \end{tikzpicture}
         \end{scriptsize}
    \end{array}\;.
\end{equation}
The corresponding magnetic quiver, obtained following the methods of \cite{Akhond:2020vhc,Akhond:2021knl} is 
\begin{equation}\label{OSp MQ Nv<N-5 Nv odd}
    \begin{array}{c}
         \begin{scriptsize}
         \begin{tikzpicture}
         \node[label=below:{1}][sof](o1){};
         \node[label=below:{2}][sp](sp1)[right of=o1]{};
         \node[label=below:{3}][so](so3)[right of=sp1]{};
         \node (dots)[right of=so3]{$\cdots$};
         \node[label=below:{$N_\textbf{v}$}][so](spn-3)[right of=dots]{};
         \node[label=below:{$N_\textbf{v}+1$}][sp](so2n-5)[right of=spn-3]{};
         \node[label=below:{$N_\textbf{v}$}][so](spn-3')[right of=so2n-5]{};
         \node (dots')[right of=spn-3']{$\cdots$};
         \node[label=below:{3}][so](so3')[right of=dots']{};
         \node[label=below:{2}][sp](sp1')[right of=so3']{};
         \node[label=below:{1}][so](o1')[right of=sp1']{};
         \node[label=left:{1}][u](u1)[above of=so2n-5]{};
         \node[label=left:{$3$}][uf](uf)[above left of=u1]{};
         \node[label=right:{$r-\frac{N_\textbf{v}-1}{2}-3$}][uf](uf')[above right of=u1]{};
         \path [draw,snake it](u1)--(uf');
         \draw(o1)--(sp1);
         \draw(sp1)--(so3);
         \draw(so3)--(dots);
         \draw(dots)--(spn-3);
         \draw(spn-3)--(so2n-5);
         \draw(so2n-5)--(spn-3');
         \draw(spn-3')--(dots');
         \draw(dots')--(so3');
         \draw(so3')--(sp1');
         \draw(sp1')--(o1');
         \draw(so2n-5)--(u1);
         \draw(u1)--(uf);
         \end{tikzpicture}
         \end{scriptsize}
    \end{array}\;.
\end{equation}
Notice that the central USp$(N_\textbf{v}+1)$ node appearing in this magnetic quiver is bad in the sense of Gaiotto and Witten \cite{Gaiotto:2008ak}. Due to this technical reason we are not able to perform an explicit computation of its Coulomb branch Hilbert series. There is however an alternative consistency check one could perform for $N_\textbf{v}=1$, by comparing the Higgs branch Hilbert series. Note that in this limit, the non-simply-laced edge in \eqref{MQ Nv<N-5} disappears and such a computation is accessible. Concretely, one has to compare the Molien-Weyl formula for the $N=2r$, $N_\textbf{v}=1$ limit of \eqref{MQ Nv<N-5}
\begin{equation}
   \oint_{|p|=1}\frac{dp}{2\pi \text{i}p}\oint_{|q|=1}\frac{dq}{2\pi \text{i}q} \frac{\PE\left[\left(\frac{p}{q}+\frac{q}{p}\right)t+(p+p^{-1})t+(r-3)(q^2+q^{-2})t+3(q+q^{-1})t\right]}{\PE\left[2t^2\right]}\;,
\end{equation}
with the Molien-Weyl formula for the $N_\textbf{v}=1$ limit of \eqref{OSp MQ Nv<N-5 Nv odd}
\begin{equation}\begin{gathered}
  \sum_{p\in\{1,-1\}}\oint_{|q|=1}\frac{dq}{2\pi\text{i}q}\oint_{|u|=1}\frac{du}{2\pi\text{i}u}(1-u^2)\times\\  \times\frac{\PE\left[(q+q^{-1})[1]_{u}t+(r-3)(q^2+q^{-2})t+3(q+q^{-1})t+(2+p+p^{-1})[1]_ut\right]}{\PE\left[\left(1+[2]_{u}\right)t^2\right]}
   \end{gathered}\;.
\end{equation}
Indeed, evaluating these integrals by computing their residues one finds an exact agreement. We collect the results for several values of $N$ in table \ref{tab:Higgs branch Nv=1}.
\begin{table}[!htb]
    \centering
    \begin{tabular}{|c|c|}\hline
         $\begin{array}{c}
              N
         \end{array}$&$\begin{array}{c}
              \HS_\mathcal{H}(t)
         \end{array}$  \\\hline
         $\begin{array}{c}
              6
         \end{array}$&$\begin{array}{c}
              \frac{1 + t + 6 t^2 + 9 t^3 + 15 t^4 + 12 t^5 + 15 t^6 + 9 t^7 + 
 6 t^8 + t^9 + t^{10}}{(1 - t)^6 (1 + t)^4 (1 + t + t^2)^3}
         \end{array}$\\\hline
         $\begin{array}{c}
              8
         \end{array}$&$\begin{array}{c}
              \frac{1 + t + 7 t^2 + 21 t^3 + 39 t^4 + 58 t^5 + 90 t^6 + 110 t^7 + 
 118 t^8 + 110 t^9 + 90 t^{10} + 58 t^{11} + 39 t^{12} + 21 t^{13} + 
 7 t^{14} + t^{15} + t^{16}}{(1 - t)^8 (1 + 2 t + 2 t^2 + t^3)^4 (1 + t + 
   t^2 + t^3 + t^4)}
         \end{array}$\\\hline
         $\begin{array}{c}
              10
         \end{array}$&$\begin{array}{c}
              \frac{1 + t + 10 t^2 + 35 t^3 + 82 t^4 + 171 t^5 + 324 t^6 + 517 t^7 + 
 740 t^8 + 961 t^9 + 1113 t^{10} + 1158 t^{11} + 1113 t^{12} +\dotsb\text{palindrome}\dotsb + t^{22}}{(1 - t)^{10} (1 + t)^4 (1 + t + t^2)^5 (1 + 
   t + t^2 + t^3 + t^4)^2}
         \end{array}$\\\hline
    \end{tabular}
    \caption{Higgs branch Hilbert series of \eqref{MQ Nv<N-5} and \eqref{OSp MQ Nv<N-5 Nv odd} for $N_\textbf{v}=1$ and various values of $N$.}
    \label{tab:Higgs branch Nv=1}
\end{table}
Finally, let us mention that for the SO(6)+1\textbf{v}, namely for the case when $N=6$ and $N_\textbf{v}=1$, there is also a unitary web description, that of SU(4)$_0$+1 \textbf{AS}, depicted in figure \ref{fig:SU(4)_0 + 1AS}. Indeed the magnetic quiver one obtains from this ordinary web diagram at infinite coupling is
\begin{equation}
    \begin{array}{c}
         \begin{scriptsize}
         \begin{tikzpicture}
         \node[label=below:{1}][u](u1){};
         \node[label=left:{1}][u](u1')[above left of=u1]{};
         \node[label=right:{1}][u](u1'')[above right of=u1]{};
         \draw(u1)--(u1');
         \draw(u1)--(u1'');
         \draw[double distance=3pt](u1')--(u1'');
         \draw(u1')--(u1'');
         \end{tikzpicture}
         \end{scriptsize}
    \end{array}\longleftrightarrow \begin{array}{c}
         \begin{scriptsize}
         \begin{tikzpicture}
         \node[label=below:{1}][u](u1){};
         \node[label=below:{1}][u](u1')[right of=u1]{};
         \node[label=above:{3}][uf](uf)[above of=u1]{};
         \node[label=above:{1}][uf](uf')[above of=u1']{};
         \draw(u1)--(u1');
         \draw(u1)--(uf);
         \draw(u1')--(uf');
         \end{tikzpicture}
         \end{scriptsize}
    \end{array}\;,
\end{equation}
where the two quivers above are related by decoupling an overall U(1), and furthermore the quiver on the right matches our proposed magnetic quiver for SO(6)+1\textbf{v} \eqref{MQ Nv<N-5} on the nose. This provides us with yet another consistency check.
\subsection{\texorpdfstring{SO$(N)$ with $N_\textbf{v}=N-4$ flavours}{TEXT}}
Let us now consider SO$(N)$ gauge theories with $N_\textbf{v}=N-4$ hypermultiplets in the vector representation. The starting configurations are the web diagrams in figure \ref{fig:brane web with flavour D7s} with $N_\textbf{v}=N-4$ flavour D7 branes. The Higgs branch directions are most easily read off, by performing a few Hanany-Witten moves which we now describe. One first moves all flavour D7 branes through, say, the left 5-brane, creating $N_\textbf{v}$ D5 branes in the process and changing the charge of the 5 brane which crosses their monodromy cut. In the SO$(2r)$ (resp. SO$(2r+1)$)  web in figure \ref{fig:brane web with flavour D7s} this results in the $(r-k-2,-1)$ (resp. $(r-k-1,-1)$) being converted into an $(r+k-2,1)$ leading, after tuning all mass parameters to zero, to the following web diagram
\begin{equation}
    \begin{array}{c}\begin{scriptsize}
    \begin{tikzpicture}
    \draw[thick](2,0)--(0,-2);
    \node[label=right:{$(r+k-2,1)$}][7brane] at (2,0){};
    \node[7brane] at (1,-1){};
    \node[label=below right:{O7$^+$}][cross] at (0,-2){};
    \draw[thick](0,-2)--(-2,-2);
    \draw[thick](-5,-2)--(-3,-2);
    \node at (-2.5,-2){$\cdots$};
    \node[7brane] at (-1,-2){};
    \node[7brane] at (-2,-2){};
    \node[7brane] at (-3,-2){};
    \node[7brane] at (-4,-2){};
    \node[7brane] at (-5,-2){};
    \node [label=below:{1}] at (-4.5,-2){};
    \node [label=below:{2}] at (-3.5,-2){};
    \node [label=below:{$N-5$}] at (-1.5,-2){};
    \node [label=below:{$N-4$}] at (-.5,-2){};
    \node[label=right:{1}] at (1.5,-.5){};
    \node[label=right:{2}] at (0.5,-1.5){};
    \end{tikzpicture}
    \end{scriptsize}\end{array}\;,
\end{equation}
from here, following a similar logic that led us to the magnetic quiver \eqref{MQ Nv<N-5} from the brane web in figure \ref{fig:SO(N)+Nv<5 web}, we obtain the following magnetic quiver
\begin{equation}
\label{eq:SON_Nv=N-4}
    \begin{array}{c}\begin{scriptsize}
    \begin{tikzpicture}
    \node[label=below:{1}][u](1){};
    \node[label=below:{2}][u](2)[right of=1]{};
    \node (dots)[right of=2]{$\cdots$};
    \node[label=below:{$N-5$}][u](2N-5)[right of=dots]{};
    \node[label=below:{$N-4$}][u](2N-4)[right of=2N-5]{};
    \node[uf][right of=2N-4]{};
    \node[label=below:{$2$}][u](2')[right of=2N-4]{};
    \node[label=below:{$1$}][u](1')[right of=2']{};
    \node[label=above:{2}][uf](uf)[above of=2']{};
    \draw(uf)--(2');
    \draw(1)--(2);
    \draw(2)--(dots);
    \draw(dots)--(2N-5);
    \draw[ double distance=1.5pt,<-](2N-5)--(2N-4);
    \draw(2N-4)--(2');
    \draw[double distance=1.5pt,->](2')--(1');
    \end{tikzpicture}
    \end{scriptsize}\end{array}\;.
\end{equation}
Notice that generically the only unbalanced node is the squircle, which suggests the Coulomb branch isometry in the generic case to be $\mathfrak{usp}(2N-8)\oplus \mathfrak{su}(2)$. The only exception to this happens for $N=4$. As we will describe in more detail below, in this case the moduli space is actually two copies of $\mathbb{C}^2/\mathbb{Z}_2$ and hence one has an $\mathfrak{su}(2)\oplus\mathfrak{su}(2)$ Coulomb branch symmetry.
Note that the number of flavours attached to the squircle is due to the O7$^+$-plane. The reason that this is 2, rather than 4 as in the previous cases, is purely fixed by trial and error and by the requirement to match the computation on the OSp side as outlined below. For now let us conjecture this quiver and slowly build up the evidence in its support. 

We can compare the Coulomb branch of \eqref{eq:SON_Nv=N-4}  for $N=2r$, with the OSp magnetic quiver for SO($2r$) with $N_\textbf{v}=2r-4$, obtained from the brane web of this theory which uses O5$^+$-planes:
\begin{equation}
    \begin{array}{c}
         \begin{scriptsize}
         \begin{tikzpicture}
         \draw[thick,dashed](0,0)--(1,0);
         \draw[thick](1,0)--(3,0);
         \draw[thick](4,0)--(9,0);
         \draw[thick](12,0)--(10,0);
         \draw[thick](6.5,0)--(6.5,2);
         \draw[thick,dashed](13,0)--(12,0);
         \node[7brane]at(6.5,1){};
         \node[7brane]at(6.5,2){};
         \node[7brane]at(1,0){};
         \node[7brane]at(2,0){};
         \node[7brane]at(3,0){};
         \node[7brane]at(4,0){};
         \node[7brane]at(5,0){};
         \node[7brane]at(9,0){};
         \node[7brane]at(8,0){};
         \node[7brane]at(10,0){};
         \node[7brane]at(11,0){};
         \node[7brane]at(12,0){};
         \node at (3.5,0){$\cdots$};
         \node at (9.5,0){$\cdots$};
         \node[label=right:{1}]at (6.5,1.5){};
         \node[label=right:{2}]at (6.5,.5){};
         \node[label=below:{O5$^+$}] at (.5,0){};
         \node[label=below:{O5$^+$}] at (12.5,0){};
         \node[label=below:{$1$}] at (1.5,0){};
         \node[label=below:{$1$}] at (2.5,0){};
         \node[label=below:{$r-2$}] at (4.5,0){};
         \node[label=below:{$r-2$}] at (5.75,0){};
         \node[label=below:{$r-2$}] at (7.25,0){};
         \node[label=below:{$r-2$}] at (8.5,0){};
         \node[label=below:{$1$}] at (10.5,0){};
         \node[label=below:{$1$}] at (11.5,0){};
         \end{tikzpicture}
         \end{scriptsize}
    \end{array}\;.
\end{equation}
The magnetic quiver one obtains from this web diagram is given by
\begin{equation}\label{Nv=N-4 OSp MQ}
    \begin{array}{c}
         \begin{scriptsize}
         \begin{tikzpicture}
         \node[label=below:{1}][so](o1){};
         \node[label=below:{2}][sp](sp1)[right of=o1]{};
         \node[label=below:{3}][so](so3)[right of=sp1]{};
         \node (dots)[right of=so3]{$\cdots$};
         \node[label=below:{$2r-4$}][sp](spn-3)[right of=dots]{};
         \node[label=below:{$2r-3$}][so](so2n-5)[right of=spn-3]{};
         \node[label=below:{$2r-4$}][sp](spn-3')[right of=so2n-5]{};
         \node (dots')[right of=spn-3']{$\cdots$};
         \node[label=below:{3}][so](so3')[right of=dots']{};
         \node[label=below:{2}][sp](sp1')[right of=so3']{};
         \node[label=below:{1}][so](o1')[right of=sp1']{};
         \node[label=left:{2}][sp](u1)[above of=so2n-5]{};
         \node[label=right:{3}][sof](uf)[right of=u1]{};
         \node[label=left:{1}][u](uu1)[above of=u1]{};
         \draw(u1)--(uu1);
         \draw(o1)--(sp1);
         \draw(sp1)--(so3);
         \draw(so3)--(dots);
         \draw(dots)--(spn-3);
         \draw(spn-3)--(so2n-5);
         \draw(so2n-5)--(spn-3');
         \draw(spn-3')--(dots');
         \draw(dots')--(so3');
         \draw(so3')--(sp1');
         \draw(sp1')--(o1');
         \draw(so2n-5)--(u1);
         \draw(u1)--(uf);
         \end{tikzpicture}
         \end{scriptsize}
    \end{array}\;.
\end{equation}
We computed the Coulomb branch HS for the two quivers in \eqref{eq:SON_Nv=N-4} and \eqref{Nv=N-4 OSp MQ} and found agreement. We provide the results for small values of $N$ below.
\subsubsection{\texorpdfstring{$N$=4}{TEXT}}
For $N=4$ this theory is the pure SO(4) so it can be engineered as two copies of pure SU$(2)$ without the need of an O7 plane. From this an analysis the ordinary magnetic quiver expected, gives two copies of $\mathbb{C}^2/\mathbb{Z}_2$ \cite{Akhond:2021knl}. The CB HS that we expect is then
\begin{equation}
    \HS\rvert_{N=4}=\frac{1+2t^2+t^4}{(1-t^2)^4}
\end{equation}
Indeed, we computed the CB HS of (\ref{eq:SON_Nv=N-4}) for $N=4$ and we find a match.
\subsubsection{\texorpdfstring{$N$=5}{TEXT}}
\begin{figure}[!htb]
    \centering
    \begin{scriptsize}
\begin{tikzpicture}
        \draw[thick](2,-2)--(3.5,-2);
        \draw[thick](2,-2)--(2,-.5);
        \draw[thick](3,-1)--(3,-1.5);
        \draw[thick](3,-1)--(2.5,-1);
        \node[7brane][label=above:{(1,-1)}] at (1.4,0){};
        \node[7brane][label=above:{(1,1)}] at (3.95,0.05){};
        \node[7brane][label=below:{(1,1)}] at (1.55,-2.55){};
        \node[7brane][label=below:{(1,-1)}] at (4.15,-2.55){};
        \draw[thick](2,-2)--(1.5,-2.5);
        \draw[thick](2.1,-2.1)--(1.6,-2.6);
        
        \draw[thick](3,-1)--(3.4,-.6);

        \draw[thick](2.5,-1.5)--(3,-1.5);
        \draw[thick](2.5,-1.5)--(2.5,-1);
        \draw[thick](3.5,-.5)--(3.5,-2);
        \draw[thick](4.1,-2.6)--(3.5,-2);
        \draw[thick](4.2,-2.5)--(3.6,-1.9);
        \draw[thick](3,-1.5)--(3.4,-1.9);
        \draw[thick](3.5,-.5)--(2,-.5);
        \draw[thick](2.5,-1)--(2.1,-.6);
        \draw[thick](1.5,0)--(2,-.5);
        \draw[thick](1.4,-.1)--(1.9,-.6);
        \draw[thick](2.5,-1.5)--(2.1,-1.9);
        \draw[thick](3.5,-.5)--(4,0);
        \draw[thick](3.4,-.4)--(3.9,0.1);
\end{tikzpicture}
    \end{scriptsize}
\hspace{.5cm}    
\begin{scriptsize}
\begin{tikzpicture}[scale=.75]
\draw[thick](-1,1)--(1,3);
\draw[thick](1,1)--(-1,3);
\node[7brane][label=above:{(1,-1)}] at (-1,3){};
\node[7brane] at (1,1){};
\node[7brane] at (-1,1){};
\node[7brane][label=above:{(1,1)}] at (1,3){};
\node[label=right:{2}] at (.5,1.5){};
\node[label=left:{2}] at (-.5,2.5){};
\node[label=above:{2}] at (.5,2.5){};
\node[label=below:{2}] at (-.5,1.5){};
\node at (0,-.5){};
\end{tikzpicture}
\end{scriptsize}
    \caption{Brane web for USp(4)+1 \textbf{AS}}
    \label{fig:USp(4)+1AS web}
\end{figure}
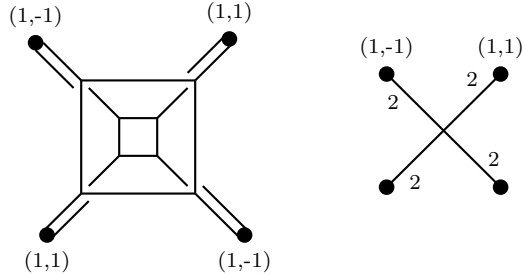
When $N=5$, the 5d theory in question is SO(5)+1\textbf{v}, which is the same as USp$(4)$+1\textbf{AS}. The latter theory can be engineered using an ordinary brane web which we depict in figure \ref{fig:USp(4)+1AS web}. The magnetic quiver that one reads from the brane web at infinite coupling is
\begin{equation}
    \begin{array}{c}
         \begin{scriptsize}
         \begin{tikzpicture}
         \node[label=below:{2}][u](u2){};
         \node[label=below:{2}][u](u2')[right of=u2]{};
         \draw[double distance=2pt](u2)--(u2');
         \end{tikzpicture}
         \end{scriptsize}
    \end{array}\;.
\end{equation}
Note that this quiver is bad in the sense of Gaiotto and Witten and so we cannot compute its Coulomb branch. However, we can make a prediction about what its Coulomb branch Hilbert series would be, utilising the non-simply-laced description. There is actually a third quiver, which should also have the same CB HS, as first discussed in \cite{Akhond:2020vhc}. This is the OSp magnetic quiver for SO(5)+1\textbf{v}, obtained from the brane web using $\widetilde{\text{O5}^+}$-plane. In particular, we claim the following three quivers to have identical Coulomb branch HS 
\begin{equation}\label{triality 2 su(2) instantons}
\begin{array}{c}\begin{scriptsize}
    \begin{tikzpicture}
    \node[label=below:{$1$}][u](2N-4)[right of=2N-5]{};
    \node[uf][right of=2N-4]{};
    \node[label=below:{$2$}][u](2')[right of=2N-4]{};
    \node[label=below:{$1$}][u](1')[right of=2']{};
    \node[label=above:{2}][uf](uf)[above of=2']{};
    \draw(uf)--(2');
    \draw(2N-4)--(2');
    \draw[double distance=1.5pt,->](2')--(1');
    \end{tikzpicture}
    \end{scriptsize}\end{array}\longleftrightarrow
    \begin{array}{c}
         \begin{scriptsize}
         \begin{tikzpicture}
         \node[label=below:{1}][u](u1){};
         \node[label=below:{2}][sp](sp2)[right of=u1]{};
         \node[label=below:{3}][so](so3)[right of=sp2]{};
         \node[label=above:{3}][sof](so3f)[above of=sp2]{};
         \draw(u1)--(sp2);
         \draw(sp2)--(so3);
         \draw(sp2)--(so3f);
         \end{tikzpicture}
         \end{scriptsize}
    \end{array}\longleftrightarrow
    \begin{array}{c}
         \begin{scriptsize}
         \begin{tikzpicture}
         \node[label=below:{2}][u](u2){};
         \node[label=above:{2}][u](u2')[above of=u2]{};
         \draw[double distance=2pt](u2)--(u2');
         \end{tikzpicture}
         \end{scriptsize}
    \end{array}
\end{equation}
The HS one computes for the non-simply-laced quiver in \eqref{triality 2 su(2) instantons} is
\begin{equation}
    \HS\rvert_{N=5}= \frac{1+t+3t^2+6t^3+8t^4+6t^5+8t^6+6t^7+3t^8+t^9+t^{10}}{(1-t)^6(1+t)^4(1+t+t^2)^3}\;.
\end{equation}
We recognise this as the HS for the reduced moduli space of two-SU(2) instantons on $\mathbb{C}^2$ \cite{Hanany:2012dm}. 
\subsubsection{\texorpdfstring{$N$=6}{TEXT}}
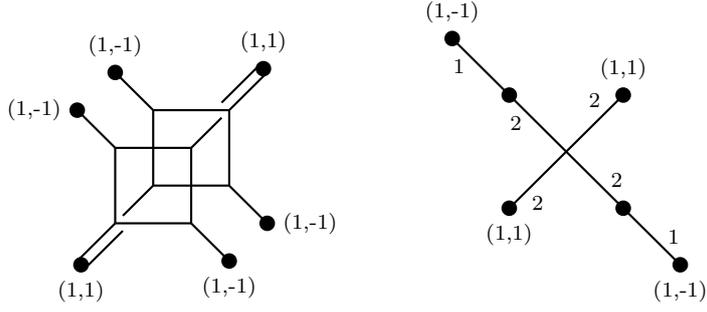
\begin{figure}
    \centering
    \begin{scriptsize}
\begin{tikzpicture}
        \draw[thick](2,-2)--(3,-2);
        \draw[thick](2,-2)--(2,-1);
        \draw[thick](3,-1)--(3,-2);
        \draw[thick](3,-1)--(2,-1);
        \node[7brane][label=left:{(1,-1)}] at (1.5,-0.5){};
        \node[7brane][label=above:{(1,-1)}] at (2,0){};
        \node[7brane][label=above:{(1,1)}] at (3.95,0.05){};
        \node[7brane][label=right:{(1,-1)}] at (4,-2){};
        \node[7brane][label=below:{(1,-1)}] at (3.5,-2.5){};
        \node[7brane][label=below:{(1,1)}] at (1.55,-2.55){};
        \draw[thick](3,-2)--(3.5,-2.5);
        \draw[thick](2,-2)--(1.5,-2.5);
        \draw[thick](2.1,-2.1)--(1.6,-2.6);
        \draw[thick](2,-1)--(1.5,-.5);
        \draw[thick](3,-1)--(3.4,-.6);

        \draw[thick](2.5,-1.5)--(3.5,-1.5);
        \draw[thick](2.5,-1.5)--(2.5,-.5);
        \draw[thick](3.5,-.5)--(3.5,-1.5);
        \draw[thick](3.5,-.5)--(2.5,-.5);
        
        \draw[thick](3.5,-1.5)--(4,-2);
        \draw[thick](2.5,-1.5)--(2.1,-1.9);
        \draw[thick](2.5,-.5)--(2,0);
        \draw[thick](3.5,-.5)--(4,0);
        \draw[thick](3.4,-.4)--(3.9,0.1);
\end{tikzpicture}
    \end{scriptsize}
\hspace{.5cm}    
\begin{scriptsize}
\begin{tikzpicture}[scale=.75]
\draw[thick](-1,-1)--(1,1);
\draw[thick](2,-2)--(-2,2);
\node[7brane][label=above:{(1,-1)}] at (-2,2){};
\node[7brane] at (-1,1){};
\node[7brane] at (1,-1){};
\node[7brane][label=below:{(1,-1)}] at (2,-2){};
\node[7brane][label=below:{(1,1)}] at (-1,-1){};
\node[7brane][label=above:{(1,1)}] at (1,1){};
\node[label=right:{2}] at (.5,-.5){};
\node[label=right:{1}] at (1.5,-1.5){};
\node[label=left:{2}] at (-.5,.5){};
\node[label=left:{1}] at (-1.5,1.5){};
\node[label=above:{2}] at (.5,.5){};
\node[label=below:{2}] at (-.5,-.5){};
\end{tikzpicture}
\end{scriptsize}
    \caption{Brane web for SU(4)$_0$+2 \textbf{AS}}
    \label{fig:SU(4)+2AS web}
\end{figure}
This theory is SO$(6)$ with $2$ vectors. Since this theory is isomorphic to SU$(4)_0$+ 2 \textbf{AS}, we can extract a unitary magnetic quiver from a brane system that doesn't involve O7 planes. The brane web for SU(4)$_0$ + 2 \textbf{AS} is depicted in figure \ref{fig:SU(4)+2AS web}. From this brane web one can readily obtain the following magnetic quiver
\begin{equation}
\label{MQ SU(4)+2AS}
\begin{array}{c}
     \begin{scriptsize}
     \begin{tikzpicture}
     \node[label=below:{1}][u](1){};
     \node[label=below:{2}][u](2)[right of=1]{};
     \node[label=below:{1}][u](1')[right of=2]{};
     \node[label=above:{2}][u](11)[above of=2]{};
     \draw(1)--(2);
     \draw(2)--(1');
     \draw[double distance=2pt](2)--(11);
     \end{tikzpicture}
     \end{scriptsize}
\end{array}
\end{equation}
We therefore claim that the following three quivers should have the same Coulomb branch
\begin{equation}
    \begin{array}{c}\begin{scriptsize}
    \begin{tikzpicture}
    \node[label=below:{$1$}][u](2N-5)[right of=dots]{};
    \node[label=below:{$2$}][u](2N-4)[right of=2N-5]{};
    \node[uf][right of=2N-4]{};
    \node[label=below:{$2$}][u](2')[right of=2N-4]{};
    \node[label=below:{$1$}][u](1')[right of=2']{};
    \node[label=above:{2}][uf](uf)[above of=2']{};
    \draw(uf)--(2');
    \draw[ double distance=1.5pt,<-](2N-5)--(2N-4);
    \draw(2N-4)--(2');
    \draw[double distance=1.5pt,->](2')--(1');
    \end{tikzpicture}
    \end{scriptsize}\end{array}\longleftrightarrow \begin{array}{c}
         \begin{scriptsize}
         \begin{tikzpicture}
         \node[label=below:{1}][so](o1){};
         \node[label=below:{2}][sp](sp1)[right of=o1]{};
         \node[label=below:{$3$}][so](so2n-5)[right of=sp1]{};
         \node[label=below:{2}][sp](sp1')[right of=so2n-5]{};
         \node[label=below:{1}][so](o1')[right of=sp1']{};
         \node[label=left:{2}][sp](u1)[above of=so2n-5]{};
         \node[label=right:{3}][sof](uf)[right of=u1]{};
         \node[label=left:{1}][u](uu1)[above of=u1]{};
         \draw(u1)--(uu1);
         \draw(o1)--(sp1);
         \draw(sp1)--(so2n-5);
         \draw(sp1')--(so2n-5);
         \draw(sp1')--(o1');
         \draw(so2n-5)--(u1);
         \draw(u1)--(uf);
         \end{tikzpicture}
         \end{scriptsize}
    \end{array}\longleftrightarrow
    \begin{array}{c}
     \begin{scriptsize}
     \begin{tikzpicture}
     \node[label=below:{1}][u](1){};
     \node[label=below:{2}][u](2)[right of=1]{};
     \node[label=below:{1}][u](1')[right of=2]{};
     \node[label=above:{2}][u](11)[above of=2]{};
     \draw(1)--(2);
     \draw(2)--(1');
     \draw[double distance=2pt](2)--(11);
     \end{tikzpicture}
     \end{scriptsize}
\end{array}\;.
\end{equation}
The expected CB HS for the magnetic quiver is then
\begin{equation}
        \HS\rvert_{N=6}=\frac{1+8t^2+40t^4+107t^6+199t^8+234t^{10}+199t^{12}+\dotsb\text{palindrome}\dotsb+t^{20}}{(1-t^2)^{10}(1+t^2)^5}
\end{equation}
We computed the CB HS of (\ref{eq:SON_Nv=N-4}) for $N=6$ and we find a match.
\subsection{\texorpdfstring{SO$(N)$ with $N_\textbf{v}=N-3$}{TEXT}}
We now analyse the final possible family of theories, SO($N$) gauge theories with $N_\textbf{v}=N-3$ hypermultiplets in the vector representation. This is the highest possible number of vector matter that has a UV completion in 5d. The web diagram for the fixed point limit can be obtained by adding D7 branes to the pure gauge theory cases as in figure \ref{fig:brane web with flavour D7s}. To read off the Higgs branch directions we first perform the following sequence of Hanany-Witten moves. First, one has to move all $2r-3$ (resp. $2r-2$)  D7s to, say, the left of the $(r-k-2,-1)$ (resp. $(r-k-1,-1)$) 5-brane in the web for the SO$(2r)$ (resp. SO$(2r+1)$) theory in figure \ref{fig:brane web with flavour D7s}. In doing so, the $(r-k-2,-1)$ and $(r-k-1,-1)$ 5-branes are both converted into an $(r+k-1,1)$. Next we pull the $(r+k-1,1)$ 5-brane through the $(r+k-2,1)$ 5-brane, which creates an additional $(r+k-1,1)$ 5-brane ending on the $(r+k-1,1)$ 7-brane in addition to converting the $(r+k-2,1)$ 5-brane into a D5. Setting all mass parameters to zero and separating the D7s to make the Higgs branch directions manifest, we arrive at    
\begin{equation}
    \begin{scriptsize}
    \begin{tikzpicture}
    \draw[thick](1.5,-1)--(0,-2);
    \node[label=right:{$(r+k-1,1)$}][7brane] at (1.5,-1){};
    \node[label=below right:{O7$^+$}][cross] at (0,-2){};
    \draw[thick](0,-2)--(-2,-2);
    \draw[thick](-5,-2)--(-3,-2);
    \node at (-2.5,-2){$\cdots$};
    \node[7brane] at (-1,-2){};
    \node[7brane] at (-2,-2){};
    \node[7brane] at (-3,-2){};
    \node[7brane] at (-4,-2){};
    \node[7brane] at (-5,-2){};
    \node [label=below:{1}] at (-4.5,-2){};
    \node [label=below:{2}] at (-3.5,-2){};
    \node [label=below:{$N-3$}] at (-1.5,-2){};
    \node [label=below:{$N-2$}] at (-.5,-2){};
    \node[label=right:{2}] at (0,-1.5){};
    \end{tikzpicture}
    \end{scriptsize}\;.
\end{equation}
The magnetic quiver we propose in this case is given by
\begin{equation}
\label{eq:SO2N_Nf=2N-3}
\begin{array}{c}    \begin{scriptsize}
    \begin{tikzpicture}
    \node[label=below:{1}][u](1){};
    \node[label=below:{2}][u](2)[right of=1]{};
    \node (dots)[right of=2]{$\cdots$};
    \node[label=below:{$N-3$}][u](2N-5)[right of=dots]{};
    \node[label=below:{$N-2$}][u](2N-4)[right of=2N-5]{};
    \node[uf][right of=2N-4]{};
    \node[label=below:{$2$}][u](2')[right of=2N-4]{};
    \node[label=above:{2}][uf](uf)[above of=2']{};
    \draw(2')--(uf);
    \draw(1)--(2);
    \draw(2)--(dots);
    \draw(dots)--(2N-5);
    \draw[ double distance=1.5pt,<-](2N-5)--(2N-4);
    \draw(2N-4)--(2');
    \end{tikzpicture}
    \end{scriptsize}\end{array}\;,
\end{equation}
where all the gauge nodes and hypermultiplet contributions are read off following a discussion analogous to that outlined in the previous cases. We stress that the rank of the flavour node attached to the squircle is fixed by the requirement that the CB HS matches that of the OSp quiver below. It would be interesting to understand the number of flavours more clearly. Except for the squircle, all other nodes in this quiver all balanced, with the balanced subquiver forming the Dynkin diagram of $C_{N-2}$, implying that the Coulomb branch isometry is generically $\mathfrak{usp}(2N-4)$.

Since SO($N$) theory with $N_\textbf{v}=N-3$ hypers in the vector representation can also be engineered using an O5-plane, we can provide a consistency check of the magnetic quiver \eqref{eq:SO2N_Nf=2N-3}. The infinite coupling limit of the web with O5-plane in the case where $N=2r$ is given by
\begin{equation}
    \begin{array}{c}
         \begin{scriptsize}
         \begin{tikzpicture}
         \draw[thick,dashed](0,0)--(1,0);
         \draw[thick](1,0)--(3,0);
         \draw[thick](4,0)--(9,0);
         \draw[thick](12,0)--(10,0);
         \draw[thick](6.5,0)--(6.5,1);
         \draw[thick,dashed](13,0)--(12,0);
         \node[7brane]at(6.5,1){};
         \node[7brane]at(1,0){};
         \node[7brane]at(2,0){};
         \node[7brane]at(3,0){};
         \node[7brane]at(4,0){};
         \node[7brane]at(5,0){};
         \node[7brane]at(9,0){};
         \node[7brane]at(8,0){};
         \node[7brane]at(10,0){};
         \node[7brane]at(11,0){};
         \node[7brane]at(12,0){};
         \node at (3.5,0){$\cdots$};
         \node at (9.5,0){$\cdots$};
         \node[label=right:{2}]at (6.5,.5){};
         \node[label=below:{O5$^+$}] at (.5,0){};
         \node[label=below:{O5$^+$}] at (12.5,0){};
         \node[label=below:{$1$}] at (1.5,0){};
         \node[label=below:{$1$}] at (2.5,0){};
         \node[label=below:{$r-1$}] at (4.5,0){};
         \node[label=below:{$r-1$}] at (5.75,0){};
         \node[label=below:{$r-1$}] at (7.25,0){};
         \node[label=below:{$r-1$}] at (8.5,0){};
         \node[label=below:{$1$}] at (10.5,0){};
         \node[label=below:{$1$}] at (11.5,0){};
         \end{tikzpicture}
         \end{scriptsize}
    \end{array}\;,
\end{equation}
from which one can obtain the following OSp magnetic quiver
\begin{equation}
    \begin{array}{c}
         \begin{scriptsize}
         \begin{tikzpicture}
         \node[label=below:{1}][so](o1){};
         \node[label=below:{2}][sp](sp1)[right of=o1]{};
         \node[label=below:{3}][so](so3)[right of=sp1]{};
         \node (dots)[right of=so3]{$\cdots$};
         \node[label=below:{$2r-2$}][sp](spn-3)[right of=dots]{};
         \node[label=below:{$2r-1$}][so](so2n-5)[right of=spn-3]{};
         \node[label=below:{$2r-2$}][sp](spn-3')[right of=so2n-5]{};
         \node (dots')[right of=spn-3']{$\cdots$};
         \node[label=below:{3}][so](so3')[right of=dots']{};
         \node[label=below:{2}][sp](sp1')[right of=so3']{};
         \node[label=below:{1}][so](o1')[right of=sp1']{};
         \node[label=left:{2}][sp](u1)[above of=so2n-5]{};
         \node[label=left:{3}][sof](uf)[above of=u1]{};
         \draw(o1)--(sp1);
         \draw(sp1)--(so3);
         \draw(so3)--(dots);
         \draw(dots)--(spn-3);
         \draw(spn-3)--(so2n-5);
         \draw(so2n-5)--(spn-3');
         \draw(spn-3')--(dots');
         \draw(dots')--(so3');
         \draw(so3')--(sp1');
         \draw(sp1')--(o1');
         \draw(so2n-5)--(u1);
         \draw(u1)--(uf);
         \end{tikzpicture}
         \end{scriptsize}
    \end{array}\;.
\end{equation}
The computation of the CB HS of this OSp quiver is a close analog of \eqref{HS OSp Nv<N-5}, so let us skip the details and quote the final results
\begin{equation}
    \HS\rvert_{r=3}=1 + 36 t^2 + 681 t^4 + 8688 t^6 + 83376 t^8 + 640695 t^{10} + 
 4110730 t^{12} +  \mathcal{O}\left(t^{13}\right)
\end{equation}

\subsubsection{$N=4$}

For the $N=4$ case this theory is SO(4) with one vector, which can be realized without an orientifold plane as SU(2)$_0\times$ SU(2)$_0$ with a bifundamental. The magnetic quiver for this theory is 
\begin{equation}
\label{mirror of U(2) with 4 hypers}
\begin{array}{c}
     \begin{scriptsize}
     \begin{tikzpicture}
     \node[label=below:{1}][u](1){};
     \node[label=below:{2}][u](2)[right of=1]{};
     \node[label=below:{1}][u](1')[right of=2]{};
     \node[label=above:{1}][u](11)[above of=2]{};
     \draw(1)--(2);
     \draw(2)--(1');
     \draw[double distance=2pt](2)--(11);
     \end{tikzpicture}
     \end{scriptsize}
\end{array}
\end{equation}
Which leads us to claim that the following three quivers should have identical Coulomb branches
\begin{equation}\label{triality N-3}
    \begin{array}{c}    \begin{scriptsize}
    \begin{tikzpicture}
    \node[label=below:{$1$}][u](2N-5)[right of=dots]{};
    \node[label=below:{$2$}][u](2N-4)[right of=2N-5]{};
    \node[uf][right of=2N-4]{};
    \node[label=below:{$2$}][u](2')[right of=2N-4]{};
    \node[label=above:{2}][uf](uf)[above of=2']{};
    \draw(2')--(uf);
    \draw[ double distance=1.5pt,<-](2N-5)--(2N-4);
    \draw(2N-4)--(2');
    \end{tikzpicture}
    \end{scriptsize}\end{array}\longleftrightarrow \begin{array}{c}
         \begin{scriptsize}
         \begin{tikzpicture}
         \node[label=below:{1}][so](o1){};
         \node[label=below:{2}][sp](sp1)[right of=o1]{};
         \node[label=below:{$3$}][so](so2n-5)[right of=sp1]{};
         \node[label=below:{2}][sp](sp1')[right of=so2n-5]{};
         \node[label=below:{1}][so](o1')[right of=sp1']{};
         \node[label=left:{2}][sp](u1)[above of=so2n-5]{};
         \node[label=left:{3}][sof](uf)[above of=u1]{};
         \draw(o1)--(sp1);
         \draw(sp1)--(so2n-5);
         \draw(sp1')--(so2n-5);
         \draw(sp1')--(o1');
         \draw(so2n-5)--(u1);
         \draw(u1)--(uf);
         \end{tikzpicture}
         \end{scriptsize}
    \end{array}\longleftrightarrow\begin{array}{c}
     \begin{scriptsize}
     \begin{tikzpicture}
     \node[label=below:{1}][u](1){};
     \node[label=below:{2}][u](2)[right of=1]{};
     \node[label=below:{1}][u](1')[right of=2]{};
     \node[label=above:{1}][u](11)[above of=2]{};
     \draw(1)--(2);
     \draw(2)--(1');
     \draw[double distance=2pt](2)--(11);
     \end{tikzpicture}
     \end{scriptsize}
\end{array}\;.
\end{equation}
The unitary simply-laced quiver \eqref{mirror of U(2) with 4 hypers} is the 3d  mirror dual of U$(2)$ with $N_f=4$ fundamental hypermultiplets. The refined Hilbert series for the CB of \eqref{mirror of U(2) with 4 hypers} reads
\begin{equation}
\HS(t;x)=\sum_{n_1,n_2}\chi_{[n_1,2n_2,n_1]}(x)t^{n_1+2n_2}\;.
\end{equation}
We computed the unrefined HS for the CB of the non-simply laced quiver of (\ref{eq:SO2N_Nf=2N-3}) finding agreement of the results. In fact both the CB HS and the HB HS of the OSp quiver in \eqref{triality N-3} and the unitary simply-laced quiver \eqref{mirror of U(2) with 4 hypers} were computed and found to agree in \cite{Akhond:2020vhc}. We see that there is actually a third description of the Coulomb branch of these theories in terms of a non-simply-laced quiver.

%--------- Section 4 --------------
%\input{}
%--------- Conclusion -------------
\section{Discussion}\label{discussion}
In this paper we studied the Higgs branch of the UV fixed point limit of 5d $\mathcal{N}=1$ gauge theories with gauge group SO$(N)$ and matter in the vector representation. Our main approach was to use the brane web description of these theories which make use of an O7$^+$-plane. We used the brane configuration to conjecture the corresponding magnetic quivers, which were subsequently used to compute the Coulomb branch Hilbert series of the magnetic quivers. We then verified these computations by matching against OSp magnetic quivers for the same 5d theories, derived from brane webs with O5-planes. In the following we summarise some potential avenues for future explorations which are made possible due to our results.

In the current work, we restricted ourselves to 5d theories which flow to a single gauge node in the IR, an interesting immediate question one may ask, is the magnetic quivers for the fixed point limit of 5d quiver gauge theories which make use of an O7$^+$-plane. One may add another layer of complexity by considering brane webs with both O5 and O7-planes \cite{Hayashi:2015vhy}, which is likely going to lead to non-simply-laced orthosymplectic magnetic quivers. We made use of the Hilbert series computations, only as a way to verify our conjectured non-simply-laced magnetic quivers. But the Hilbert series, when refined, contains information that could be used to illuminate dynamical phenomena, for instance by comparing with the finite coupling results. We hope to report on some of these issues in future work.

We found several interesting families of non-simply-laced unitary quivers whose moduli spaces of vacua, we conjecture, are isomorphic to orthosymplectic magnetic quivers, obtained from brane webs with O5-planes. These conjectures are difficult to fully justify rigorously, as typically, either the Higgs or Coulomb branch Hilbert series computation is beyond known methods on one or both sides of the correspondence. On the other hand, whenever a Hilbert series computation was successfully performed, we found a match between the two sides. It would clearly be desirable to make further progress on this front. Two concrete potential application in this direction are;
\begin{itemize}
    \item making use of the Higgs branch Hilbert series results of the OSp side as a starting point in order to find an algorithm for computing the Higgs branch Hilbert series of non-simply-laced quivers,
    \item the Coulomb branch Hilbert series for some bad OSp or simply laced unitary quivers, were conjecturally obtained via their correspondence with good non-simply-laced quivers. Further verification of these proposals would be of interest.
\end{itemize}
Finally, our results beg for a bottom-up perspective rooted in 3d physics that would make the match of the Hilbert series of the unitary non-simply-laced quivers with the OSp quivers more intuitive. 

%---------- Acknowledgements ------
\acknowledgments
It is our pleasure to thank Siddharth Dwivedi, Hirotaka Hayashi, Sung-Soo Kim and Futoshi Yagi for useful discussions as well as collaboration on closely related topics. We are expecially thankful to Futoshi Yagi, for spotting an error in a previous version of the draft. We also wish to express our gratitude to the organisers and speakers of the QFT and Geometry summer school, as these lectures helped us develope a more broad perspective on the subject of higher dimensional SCFTs. F.C. is supported by STFC consolidated grant ST/T000708/1. MA is partially funded by an STFC consolidated grant ST/S505778/1. 
\clearpage
%--------- Appendix --------------
\appendix

%====================================================================
%%%%%%%%%%%%%%%%%%%%%%%%%%%%%%%%%%%%%%%%%%%%%%%%%%
\bibliographystyle{JHEP}

\newpage
\bibliography{ref}
\end{document}